\documentclass{article}
\usepackage{graphicx}
\usepackage{amsmath}
\begin{document}

\begin{center}
{\Large{\bf Quantum  Imaging}} \\ \vspace*{8mm} 
Yanhua Shih \\
Department of Physics \\
University of Maryland, Baltimore County, \\ Baltimore, MD 21250 
\end{center}

\vspace{1mm}

\begin{center}
\parbox{13.75cm}{{\textbf{Abstract}}: One of the most surprising consequences 
of quantum mechanics is
the entanglement of two or more distant particles.
Although questions regarding fundamental issues of quantum theory
still exist, quantum entanglement has started to play important
roles in practical engineering applications.  Quantum imaging is
one of these exciting areas.  Quantum imaging has 
demonstrated two peculiar features: 
(1) reproducing ``ghost" images in a ``nonlocal" manner, and 
(2) enhancing the spatial resolution of imaging beyond the diffraction limit.  
In this article, we start with the review of classical imaging
to establish the basic concepts and formalisms of imaging.  
We then analyze two-photon imaging with particular emphasis 
on the physics of spatial resolution enhancement and the ``ghost" 
imaging phenomenon.  
}
\end{center}

\vspace{2mm}

\section{Introduction}

\hspace{6.5mm}Quantum mechanics remarkably allows for the entanglement of 
two or more distant particles.  In a maximally entangled two-particle system, the 
value of an observable for each particle is completely undefined while the 
correlation between the two can be uniquely determined, \emph{despite the 
distance between them}.  In other words, each of the two
subsystems may have {\em completely random values} or {\em all
possible values} for some physical observable in the course of
their propagation, but the correlations of the
two subsystems are determined uniquely and with certainty through 
the measurement of joint-detection events \cite{EPR}.
Although questions regarding fundamental issues of quantum theory
still exist, quantum entanglement has started to play important
roles in practical engineering applications.  Quantum imaging is
one of these exciting areas.  

The first quantum imaging experiment was demonstrated by Pittman \emph{et al}. 
in 1995 \cite{GhostImage}, inspired by the theory of Klyshko \cite{KlyshkoImg}.  
The experiment was immediately named ``ghost imaging" 
due to its surprising nonlocal feature.  The important physics demonstrated 
in that experiment, nevertheless, may not be the ``ghost."  Indeed, 
the original purpose of the experiment was to study and to test the two-particle 
EPR correlation in position and in momentum for an entangled two-photon 
system \cite{IEEE-03}\cite{Howell}\cite{Milena}\cite{disug}.   The experiments of ghost imaging 
and ghost interference \cite{GhostInt} together stimulated the 
foundation of quantum imaging in terms of geometrical and physical optics.

Entangled multi-photon systems were later introduced to lithography for 
sub-diffraction-limitted imaging \cite{Theory}.  
In 2000, Boto \emph{et. al}. proposed a ``noon" state and approved that the entangled 
N-photon system may improve the spatial resolution of an imaging system by 
a factor of N, despite the Rayleigh diffraction limit.  The working principle of 
quantum lithography was experimentally demonstrated by D'Angelo \emph{et al}. 
in 2001 \cite{Experiment} by taking advantage of an entangled two-photon state 
of spontaneous parametric down-conversion (SPDC).  
Due to the lack of a two-photon absorber, the joint-detection measurement 
in that experiment was on the Fourier transform plane rather than on the image 
plane. The observed Fourier transform of the object function is the 
same as the one produced by classical light of wavelength $\lambda/2$. It was 
implicit in Ref.\cite{Experiment} that a second Fourier transform, by inserting a 
second lens in that experimental setup, would transfer the Fourier transform
of the object onto its image plane, thus giving an image with doubled spatial
resolution despite the Rayleigh diffraction limit.

Quantum imaging has so far demonstrated two peculiar features: 
(1) reproducing ghost images in a ``nonlocal" manner, and 
(2) enhancing the spatial resolution of imaging beyond the diffraction limit.  
Both the nonlocal behavior observed in the ghost imaging experiment and the 
apparent violation of the uncertainty principle explored in the quantum 
lithography experiment are due to the two-photon coherent effect, 
which involves the superposition of two-photon amplitudes, a nonclassical 
entity corresponding to different yet indistinguishable alternative ways of triggering 
a joint-detection event, under the framework of quantum theory of photodetection 
\cite{Glauber}.  

The nonlocal superposition of two-photon states may never be understood 
classically.   Classical attempts, however, have never stopped in the history of 
EPR studies.  Bennink \emph{et al}. demonstrated an interesting experiment in 
2002 \cite{boyd}.  In that experiment, two co-rotated laser beams produced a 
projection shadow of an object mask through coincidence measurements.   
Instead of having a superposition of a large number of two-photon probability 
{\emph{amplitudes}}, Bennink used two correlated laser beams (imagine two back 
to back lasers) to simulate each two-photon {\emph{probability}} one at a time.  
If the laser beam propagating in the object arm is blocked by 
the mask at a certain rotating angle, there would be no coincidence in
that angle and consequently the corresponding ``position" in the 
nonlocal ``image" plane, defined by the correlated laser beams which were 
propagated to a different directions.   The block-unblock of the correlated 
laser beams projected a shadow of the object mask in coincidences.  
Interestingly, this experiment has excited a number of discussions concerning 
certain historical realistic models of EPR.  Despite the fact that the 
measured correlation by Bennink \emph{et al}. is not the EPR correlation in 
momentum and, simultaneously, in position but rather a trivial 
``momentum-momentum" correlation defined by the two co-rotating laser beams 
on their Fourier transform planes, this experiment is a good example in 
distinguishing a two-photon image from a correlated projection 
shadow \cite{Milena}\cite{disug}.  

Thermal light ghost imaging was another challenge.  In 2004, Gatti \emph{et al}. 
\cite{gatti}, Wang \emph{et al}. \cite{Wang}, and Zhu \emph{et al}. \cite{Zhu}
proposed ghost imaging by replacing entangled state with chaotic thermal
radiation. A question about ghost imaging is then naturally raised: Is ghost
imaging a quantum effect if it can be simulated by ``classical" light? Thermal
light ghost imaging is based on the second-order spatial correlation of
thermal radiation. In fact, two-photon correlation of thermal radiation is not
a new observation. Hanbury-Brown and Twiss (HBT) demonstrated the intensity
spatial correlation of thermal light in 1956 \cite{hbt}. Differing from
entangled states, the maximum correlation in thermal radiation is 50\%, which
means 33\% visibility of intensity modulation at most. Nevertheless, thermal
light is a useful candidate for ghost imaging in certain applications.
Recently, a number of experiments successfully demonstrated certain
interesting features of ghost imaging by using chaotic light \cite{prl1}
\cite{prl2}\cite{gatti2}\cite{wu}.

The HBT experiment was successfully interpreted as statistical correlation of
intensity fluctuations. In HBT, the measurement is in far-field (Fourier transform 
plane). The measured two intensities have the same fluctuations while the two 
photodetectors receive the same mode and thus yield maximum correlation
\begin{equation}\label{HBT-Classical}
\langle I_{1}I_{2}\rangle=\bar{I}_{1}\bar{I}_{2}+
\langle\Delta I_{1} \Delta I_{2}\rangle.
\end{equation}
When the two photodetector receive different modes, however, the intensities
have different fluctuations, the measurement yields $\langle\Delta I_{1}
\Delta I_{2}\rangle=0$ and gives $\langle I_{1}I_{2}\rangle=\bar{I}_{1}\bar
{I}_{2}$.
One type of the HBT experiments explored the partial (50\% ) spatial
correlation $\langle I_{1}I_{2}\rangle\sim1+\delta\lbrack(\vec{\rho}_{1}%
-\vec{\rho}_{2})(\Delta\theta)]$ of the thermal radiation field, where
$\vec{\rho}_{j}$ is the transverse coordinate of the $j^{th}$ photodetector and
$\Delta\theta$ is the angular size of the source.
This result has been applied in Astronomy for measuring the angular size of stars.

Although Eq.~(\ref{HBT-Classical}) gives a reasonable explanation to the
far-field HBT phenomena, the theory of statistical correlation of intensity 
fluctuation may not work for ghost imaging: (1) this theory fails to provide 
adequate interpretation for ghost imaging of entangled states. The visibility of ghost
image of entangled two-photon state is 100\% which means a 100\% correlation,
i.e., $\langle I_{1}I_{2} \rangle\sim\delta(\vec{\rho}_{1} - \vec{\rho}_{2}/m)$, 
where $m$ is the magnification factor of imaging. If one insists on
Eq.~(\ref{HBT-Classical}), the mean intensities $\bar{I}_{1}$ and $\bar{I}%
_{2}$ must be zero. Otherwise Eq.~(\ref{HBT-Classical}) leads to non-physical
conclusions. The measurements, however, never yield zero mean values of
$\bar{I}_{1}$ and $\bar{I}_{2}$ in any circumstances. 
(2) Scarcelli \emph{et al}. recently demonstrated a lens-less ``near-field" ghost 
imaging of chaotic radiation \cite{prl2} and pointed out that the theory of 
statistical correlation of intensity fluctuation does not work for their experiment.
Differing from HBT in which the measurement is in far-field, Scarcelli's
experiment is in near-field. In near-field, for each point on the detection
plane, a point photodetector receives a large number of ($N$) modes in the
measurement. The ratio between joint-detections triggered by ``identical mode" and
joint-detections triggered by ``different modes" is $N/N^{2}$. For a large $N$, the
contributions from ``identical mode" is negligible and thus $\langle\Delta
I_{1} \Delta I_{2}\rangle=0$. Therefore, the classical idea of statistical
correlation of intensity fluctuations will not work in the multi-mode case, as
we know that different modes of chaotic light fluctuate randomly and
independently. On the other hand, Scarcelli \emph{et al}. proved a
successful alternative interpretation based on the quantum theory of two-photon 
interference.

It is interesting to see that the concept of two-photon coherence is applicable to 
both ``classical" and ``quantum" light.   Although, two-photon superposition 
is a new concept benefitted from recent research of entangled states \cite{IEEE-03}, 
the concept is not restricted to the entangled states. The concept is generally 
true and applicable to any radiation, including ``classical" thermal light.  
It is then reasonable to ask:  What is the 
difference between entangled two-photon imaging and thermal light 
two-photon imaging?   To what extent can two-photon correlation 
of chaotic radiation replace the roles of entangled states?  
Can two-photon imaging of chaotic light improve imaging 
resolution beyond the classical limit?  
Another reasonable question may trap us into an enduring debate: 
Is thermal light ghost imaging ``classical" or ``quantum"?  Is ghost 
imaging, in general, ``classical" or ``quantum"?

This article attempts to provide answers to these hotly debated
questions. To this end, we start with the review of classical imaging. 
The treatment of classical imaging  may serve
to establish the basic concept and formalism we are going to 
use in the rest of this article.  
We then analyze two-photon imaging with particular emphasis 
on the physics of spatial resolution enhancement and the ``ghost" 
imaging phenomenon.

\section{Imaging and Spatial Resolution}

\begin{figure}[htb]
\centering
    \includegraphics[width=82mm]{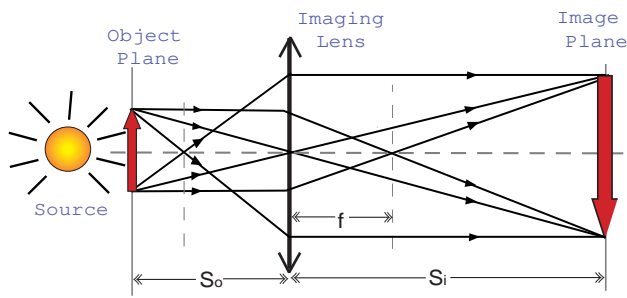}
     \parbox{14cm}{\caption{Imaging: a lens produces an
\textit{image} of an object in the plane defined by the Gaussian
thin lens equation $1/s_i+1/s_o=1/f$. The concept of an image is
based on the existence of a point-to-point relationship between
the object plane and the image plane.}\label{fig:Projection-1}}
\end{figure}

\hspace{6.5mm}The concept of imaging is well defined in classical optics. 
Fig.~\ref{fig:Projection-1} schematically illustrates a
standard imaging setup.   
A lens of finite size is used to image
the object onto an image plane which is defined by 
the ``Gaussian thin lens equation"
\begin{equation}\label{Lens-Eq}
\frac{1}{s_i}+\frac{1}{s_o}=\frac{1}{f}
\end{equation}
where $s_o$ is the distance between 
object and lens, $f$ is the
focal length of the lens and $s_i$ is the distance between lens
and image plane.  If light always follows the laws
of geometrical optics, the image plane and the object plane would
have a perfect point-to-point correspondence, which means a
perfect image of the object, either magnified or demagnified. 
Mathematically, a perfect image is the result of a convolution 
of the object distribution function $f(\vec{\rho}_{o})$ and a 
$\delta$-function.  The $\delta$-function characterize the perfect point-to-point 
relationship between the object plane and the image plane: 
\begin{eqnarray}\label{Image-Eq}
F(\vec{\rho}_i) =
\int_{obj} d\vec{\rho}_{o} \, f(\vec{\rho}_{o}) \,
\delta(\vec{\rho}_{o} + \vec{\rho}_{i} / m)
\end{eqnarray}
where $\vec{\rho}_{o}$ and $\vec{\rho}_{i}$ are 2-D vectors of the 
transverse coordinate in the object plane and the image plane,
respectively, and $m$ is the magnification factor.

Unfortunately,
light behaves like a wave. The diffraction effect turns the
point-to-point correspondence into a point-to-``spot"
relationship.  The $\delta$-function in the convolution of 
Eq.~(\ref{Image-Eq}) will be replaced by a point-spread function.  
\begin{eqnarray}\label{Image-Eq-2}
F(\vec{\rho}_i) =
\int_{obj} d\vec{\rho}_{o} \, f(\vec{\rho}_{o}) \,
somb\big{[} \frac{R}{s_o}\,\frac{\omega}{c} \big{|}\vec{\rho}_{o} + \vec{\rho}_{i} / m \big{|} \big{]} 
\end{eqnarray}
where 
$$
somb(x) = \frac{2J_1(x)}{x},
$$
and $J_1(x)$ is the first-order Bessel function, $R$ is the radius of the 
imaging lens. The finite size of the spot, which is defined by 
the point-spread function, determines the spatial resolution of
the imaging setup, and thus, limits the ability of making
demagnified images.  It is clear from Eq.~(\ref{Image-Eq-2}), the use of 
larger imaging lens and shorter wavelength light source 
will result in a narrower point-spead function.  To improve the spatial resolution, 
one of the efforts in the lithography industry is the use of shorter
wavelengths.  This effort is, however, limited to a certain level
because of the lack of lenses effectively working beyond a certain
``cutoff" wavelength.  

\begin{figure}[htb]
\centering
    \includegraphics[width=85mm]{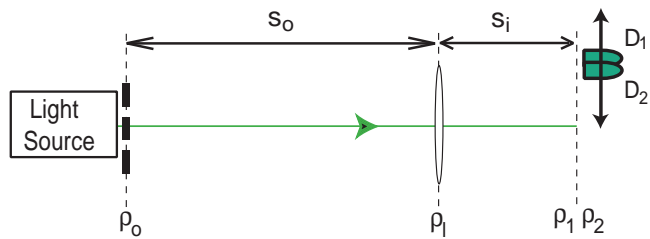}
     \parbox{14cm}{\caption{Typical imaging setup. A lens of finite size is used to produce a 
demagnified image of a object with limited spatial resolution. Replacing 
classical light with an entangled N-photon system, the spatial resolution can be 
improved by a factor of N, despite the Rayleigh diffraction limit.}\label{fig:lithography-1}}
\end{figure}

Eq.~(\ref{Image-Eq-2}) imposes a diffraction
limited spatial resolution on an imaging system while the aperture size
of the imaging system and the wavelength of the light source 
are both fixed.  This limit is fundamental in both classical
optics and in quantum mechanics.  Any violation would be considered 
as the violation of the uncertainty principle.  

Surprisingly, the use of quantum entangled states gives a different result: 
by replacing classical light sources in
Fig.~\ref{fig:lithography-1} with entangled N-photon states, the
spatial resolution of the image can be improved by a factor of N, 
despite the Rayleigh diffraction limit.  Is this a violation of the uncertainty
principle?  The answer is no!  The uncertainty relation for an entangled N-particle
system is radically different from that for N independent particles.  In terms of the 
terminology of imaging, what we have found is that the $somb(x)$ in the 
convolution of Eq.~(\ref{Image-Eq-2}) has a different
form in the case of an entangled state.  For example, an entangled two-photon 
system has 
$$
x = \frac{R}{s_o}\,\frac{2 \omega}{c} \big{|}\vec{\rho}_{o} + \vec{\rho}_{i} / m \big{|},
$$
and thus yields a twice narrower point-spread function and 
results in a doubling spatial resolution for imaging.    

\begin{figure}[htb]
\centering
    \includegraphics[width=72mm]{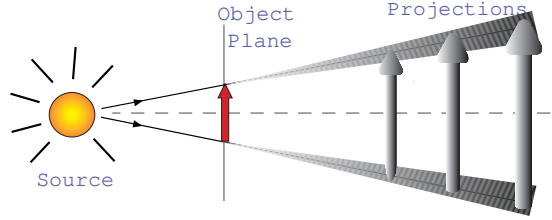}
     \parbox{14cm}{\caption{Projection: a light source
illuminates an object and no image forming system is present, no
image plane is defined, and only projections, or shadows, of the
object can be observed.}\label{fig:Projection-2}}
\end{figure}

It should be emphasized that one must not confuse a ``projection" 
with an image.  A projection is the shadow of an object, which is obviously 
different from the image of an object.  Fig.~\ref{fig:Projection-2} distinguishes
a projection shadow from an image.  The difference is more obvious 
when the projection shadow is accumulated one-ray-at-a-time, as Bennink \emph{et al}.
did in their experiment \cite{boyd}.    In a projection, the object-shadow 
correspondence is essentially a ``momentum" correspondence, which is 
defined by the propagation direction of the light rays.  This point will be
emphasized again in the section of ghost imaging.

\section{Classical Imaging}

\hspace{6.5mm}We now analyze classical imaging.
The analysis starts from the propagation of the field from
the object plane to the image plane. In classical optics, such
propagation is described by an optical transfer function
$h(\mathbf{r}-\mathbf{r}_0, t-t_0)$, which accounts for the propagation 
of all modes of the field.  In view of our quantum optics approach, we prefer 
to work with the single-mode propagator 
$g(\mathbf{k}, \mathbf{r}-\mathbf{r}_0, t-t_0)$ and its Fourier transform 
\cite{Rubin}\cite{goodman}.  
It is convenient to write the field $E(\mathbf{r}, t)$ in terms of its longitudinal ($z$)
and transverse ($\vec{\rho}$) coordinates under the Fresnel
paraxial approximation \cite{Rubin}:
\begin{eqnarray}\label{e-g}
E(\vec{\rho}, z, t) =  \int d\omega \, d\vec{\kappa} \,\, \tilde{E}(\vec{\kappa}, \omega) \,
g(\vec{\kappa}, \omega; \vec{\rho}, z) \, e^{-i\omega t}
\end{eqnarray}
where $\tilde{E}(\omega, \vec{\kappa})$ is the complex amplitude
of frequency $\omega$ and
transverse wave-vector $\vec{\kappa}$.  In Eq.~(\ref{e-g}) we have
taken $z_0=0$ and $t_0 = 0$ at the object plane as usual.
To simplify the notation, we have assumed one polarization.

Based on the experimental setup of Fig.~\ref{fig:lithography-1},
$g(\vec{\kappa}, \omega; \vec{\rho}, z)$ is found to be
\cite{Rubin}\cite{goodman}
\begin{eqnarray}\label{g-1}
& & g(\vec{\kappa}, \omega; \vec{\rho}_i, s_o+s_i) \nonumber \\
&=&  \int_{obj} d\vec{\rho}_{o} \int_{lens} d\vec{\rho}_l \, \Big\{ A(\vec{\rho}_{o}) \,
e^{i \vec{\kappa} \cdot \vec{\rho}_{o}} \Big\} \,
\Big\{\frac{-i \omega}{2 \pi c} \, 
\frac{e^{i \frac{\omega}{c} s_o} }{s_o}\, e^{i \frac{\omega}{2 c s_o} 
|\vec{\rho}_{l}-\vec{\rho}_{o}|^2} \Big\}
\,\, \Big\{ e^{-i \frac{\omega}{2 c f} \, | \vec{\rho}_l |^2} \Big\}\,
\nonumber \\ & &  \times \,  \Big\{ \frac{-i \omega}{2\pi c } \, 
\frac{e^{i \frac{\omega}{c} s_i}}{s_i}\, e^{i \frac{\omega}{2 c s_i} |\vec{\rho}_i-\vec{\rho}_l |^2 } \Big\}
\end{eqnarray}
where $\vec{\rho}_{o}$, $\vec{\rho}_l$, and $\vec{\rho}_i$ are
two-dimensional vectors defined, respectively, on the object, the lens, and the image planes.
The first curly bracket includes the aperture function $A(\vec{\rho}_{o})$ of the
object and the phase factor $e^{i \vec{\kappa} \cdot \vec{\rho}_{o}}$ contributed 
to the object plane by each transverse mode 
$\vec{\kappa}$.  The terms in the second and the fourth curly brackets describe 
free-space Fresnel propagation-diffraction from the source/object plane to the 
imaging lens, and from the imaging lens to the detection plane, respectively.
The Fresnel propagator includes a spherical wave
function $e^{i \frac{\omega}{c} (z_j -z_k)} / (z_j -z_k)$ and a Fresnel phase factor 
$e^{i \omega |\vec{\rho}_{j}-\vec{\rho}_{k}|^2 / {2 c (z_j -z_k)} } $.  
The third curly bracket adds the phase factor introduced by the imaging lens.
A brief review of Fresnel diffraction-propagation is given in Appendix A.

We now rewrite Eq.~(\ref{g-1}) into the following form
\begin{eqnarray} \label{g-2}
& & g(\vec{\kappa}, \omega; \vec{\rho}_i, z =s_o+s_i) \nonumber \\
&=&  \frac{- \omega^2}{(2 \pi c)^2 s_o s_i} \, e^{i \frac{\omega}{c} (s_o+s_i)}
\, e^{i \frac{\omega}{2 c s_i} |\vec{\rho}_i|^2} 
 \int_{obj} d\vec{\rho}_{o} \, A(\vec{\rho}_{o}) \,
e^{i \frac{\omega}{2 c s_o} |\vec{\rho}_{o}|^2} \, e^{i \vec{\kappa} \cdot \vec{\rho}_{o}}
\nonumber \\ & & \times  \int_{lens} d\vec{\rho}_l \, e^{i \frac{\omega}{2c}
[\frac{1}{s_o} + \frac{1}{s_i} - \frac{1}{f}] |\vec{\rho}_l|^2} \, e^{-i \frac{\omega}{c}
(\frac{\vec{\rho}_{o}}{s_o} + \frac{\vec{\rho}_i}{s_i}) \cdot \vec{\rho}_l}.
\end{eqnarray}
The image plane is defined by the Gaussian thin-lens equation
of Eq.~(\ref{Lens-Eq}). Hence, the second integral in
Eq.~(\ref{g-2}) simplifies and gives, for a finite sized lens of
radius $R$, the so called point-spread
function of the imaging system: $somb(x) = 2J_1(x) / x$, where $x=[\frac{R}{s_o}\,
\frac{\omega}{c} |\vec{\rho}_{o} + \rho_{i} / m|]$,
 $J_1(x)$ is the first-order Bessel function and $m=s_i/s_o$
is the magnification of the imaging system.

Substituting the result of Eqs.~(\ref{g-2}) into Eq.~(\ref{e-g}) 
enables one to obtain
the classical self-correlation of the field, or, equivalently, the intensity
on the image plane
\begin{equation}\label{i-0}
I(\vec{\rho}_i, z_i, t_i) = \langle \, E^*(\vec{\rho}_i, z_i, t_i) \, E(\vec{\rho}_i, z_i, t_i) \, \rangle
\end{equation}
where $\langle ... \rangle$ denotes an ensemble average.  
We assume monochromatic light for classical imaging 
as usual \cite{aberration}.     

\vspace{2mm}
Case (I): \textit{incoherent imaging.}
The ensemble average of
$\langle \, \tilde{E}^*(\vec{\kappa}, \omega) \,\tilde{E}(\vec{\kappa'}, \omega) \, \rangle$
yields zeros except when $\vec{\kappa}=\vec{\kappa'}$. The image is thus
\begin{eqnarray}\label{i-2}
I(\vec{\rho}_i) &\propto&
 \int d\vec{\rho}_{o} \, \big{|}A(\vec{\rho}_{o})\big{|}^2 \,
\big{|}somb[\frac{R}{s_o}\, \frac{\omega}{c} |\vec{\rho}_{o} + \frac{\vec{\rho}_{i}}{m}|] \big{|}^2.
\end{eqnarray}
An incoherent image, magnified by a factor of $m$, is
thus given by the convolution between the squared moduli of the
object aperture function and the point-spread function. The
spatial resolution of the image is thus determined by the finite
width of the $|somb|^2$-function.

\vspace{2mm}
Case (II): \textit{coherent imaging.} The coherent superposition of
the $\vec{\kappa}$ modes in both $E^*(\vec{\rho}_i, \tau)$ and
$E(\vec{\rho}_i, \tau)$ results in a wavepacket. The image, or the
intensity distribution on the image plane, is  thus
\begin{eqnarray}\label{i-1}
I(\vec{\rho}_i) \propto
\Big{|} \int_{obj} d\vec{\rho}_{o} \, A(\vec{\rho}_{o}) \,
e^{i \frac{\omega}{2 c s_o} |\vec{\rho}_{o}|^2}
somb[\frac{R}{s_o} \frac{\omega}{c} |\vec{\rho}_{o} + \frac{\vec{\rho}_{i}}{m}|] \Big{|}^2. 
\end{eqnarray}
A coherent image, magnified by a factor of $m$, is thus
given by the squared modulus of the convolution between the object
aperture function (multiplied by a Fresnel phase  factor) and
the point-spread function. 

For $s_i<s_o$ and $s_o>f$, both Eqs.~(\ref{i-2}) and (\ref{i-1})
describe a real demagnified inverted image. In both cases, a
narrower $somb$-function yields a higher spatial resolution.
Thus, the use of shorter wavelengths allows for the improving of 
the spatial resolution of an imaging system.

\section{Two-photon Imaging}

\hspace{6.5mm}To demonstrate the working principle of quantum imaging, we
replace the classical light source in Fig.~\ref{fig:lithography-1} with 
an entangled two-photon source such 
as spontaneous parametric down-conversion (SPDC) 
\cite{Klyshkobook}\cite{IEEE-03}  and
replace the ordinary film with a two-photon absorber, which is
sensitive to two-photon transition only, on the image plane. We
will show that, in the same experimental setup of
Fig.~\ref{fig:lithography-1}, an entangled two-photon system gives
rise, on a two-photon absorber, to a point-spread function twice
narrower than the one obtained in classical imaging at the same
wavelength. Then, without employing shorter wavelengths, entangled
two-photon states improve the spatial resolution of a
\emph{two-photon image} by a factor of 2.

What is special about the entangled two-photon state of SPDC?  Let
us imagine a measurement in which we place two
point-like single photon counting detectors ($D_1$ and $D_2$) on
the output surface of an SPDC crystal.
The nearly collinear signal-idler system generated by SPDC 
can be described by the two-photon state:
\begin{eqnarray}\label{State}
|\, \Psi \, \rangle
= \Psi_0 \int d \vec{\kappa}_s \,d \vec{\kappa}_i \, 
\delta(\vec{\kappa}_s + \vec{\kappa}_i )  
\int d \omega_s \, d \omega_i \, \delta(\omega_s+\omega_i-\omega_p) 
\, a^{\dag}(\vec{\kappa}_s,\omega_s) \,
a^{\dag}(\vec{\kappa}_i,\omega_i) \, |\, 0 \, \rangle
\end{eqnarray}
where $\omega_j$, $\vec{\kappa}_j$ ($j=s,i,p$), is the
frequency and the transverse wavevector of the signal, idler, and pump.
To simplify the calculation, we have assumed a CW single
frequency plane-wave pump with $\vec{\kappa}_p =0$.  The joint-detection 
probability between the point-like
detectors $D_1$ and $D_2$ located at $(\vec{\rho}_{1}, z_1)$ and $(\vec{\rho}_{2}, z_2)$, 
respectively, is calculated from the Glauber theory of photodetection \cite{Glauber}:
\begin{eqnarray}\label{G2}
G^{(2)}(\vec{\rho}_1, z_1, t_1; \vec{\rho}_2, z_2, t_2)
&=& | \langle \, 0 \, |\,
E^{(+)}_2 (\vec{\rho}_2, z_2, t_2)
E^{(+)}_1 (\vec{\rho}_1, z_1, t_1) \, |\, \Psi \, \rangle |^2 \nonumber \\
 &=&  |\, \Psi(\vec{\rho}_{1}, z_1, t_1; \vec{\rho}_{2}, z_2, t_2) \, |^2.
\end{eqnarray}
At this point we are taking $z_1=z_2=0$.  The
transverse part of the effective two-photon wavefunction $\Psi(\vec{\rho}_{1}, \vec{\rho}_{2})$
is thus calculated as:
\begin{eqnarray}\label{Wavefunction}
\Psi(\vec{\rho}_{1}, \vec{\rho}_{2})
 \simeq \Psi_0 \, \delta(\vec{\rho}_{1}-\vec{\rho}_{2}).
\end{eqnarray}
Both Eqs.~(\ref{State}) and (\ref{Wavefunction}) suggest that the
entangled signal-idler photon pair is characterized by EPR
correlation \cite{EPR} in transverse momentum and transverse position; hence,
similar to the original EPR state, we have \cite{disug}:
\begin{eqnarray}\label{eq_uncert}
& & \Delta(\vec{\kappa}_s+\vec{\kappa}_i)=0 \,\, \,\, \& \,\,\,
\Delta (\vec{\rho}_1- \vec{\rho}_2)=0 \,\, \\ \nonumber
&\text{with}&  \Delta \vec{\kappa}_s \sim \infty, \hspace{2mm}
\Delta \vec{\kappa}_i \sim \infty, \hspace{2mm}
\Delta \vec{\rho}_1 \sim \infty, \hspace{2mm} \Delta \vec{\rho}_2 \sim \infty.
\end{eqnarray}
In EPR's language, the signal and the idler may come out from any
point on the output plane of the SPDC.  However, if the
signal (idler) is found in a certain position, the idler (signal)
must be found in the same position, with certainty (100\%). Furthermore,
the signal and the idler may have any transverse momentum.
However, if a certain value and direction of the transverse
momentum of the signal (idler) is measured, the transverse
momentum of the idler (signal) will certainly have equal value and
opposite direction. 

Concerning the experimental setup of Fig.~\ref{fig:lithography-1}, in light of
Eq.~(\ref{eq_uncert}), if the object is placed very close to the output plane of the
SPDC crystal, the signal photon and the idler photon are guaranteed to come
out from the same point of the object plane ($\vec{\rho}_s = \vec{\rho}_i$) and 
to stop at the corresponding unique point on the image plane, thus forming an image 
together (i.e., a two-photon image), whether enlarged or demagnified.
In addition, the signal-idler pair propagates in a special way in which the transverse 
momenta of the pair must satisfy $\vec{\kappa}_s+\vec{\kappa}_i=0$ and thus
undergo ``two-photon diffraction."  Two-photon diffraction is
radically different from that of two independent photons and
yields a different spatial resolution of the image.  
This unique feature will be shown in the following calculation.

The fields at the detectors are the same as in the classical imaging 
case. By inserting the field operators into
$\Psi(\vec{\rho}_1, z_1, t_1; \vec{\rho}_2, z_2, t_2)$,
as defined in Eq.~(\ref{G2}), and considering the commutation
relations for the creation and the annihilation operators, we find
for the effective two-photon wavefunction in the image plane
\begin{eqnarray}\label{psi-2}
&&\Psi(\vec{\rho}_1,z_1, t_1;\vec{\rho}_2, z_2, t_2)  \\ \nonumber 
&=& \Psi_0 \int d \vec{\kappa}_s \,d \vec{\kappa}_i \,
\delta(\vec{\kappa}_s + \vec{\kappa}_i ) 
\int d \omega_s \, d \omega_i \, \delta(\omega_s+\omega_i-\omega_p)
\\ \nonumber 
&&  \times \, 
g(\vec{\kappa}_s, \omega_s; \vec{\rho}_1, z_1) \, e^{-i\omega_s t_1}  \,
g(\vec{\kappa}_i, \omega_i; \vec{\rho}_2, z_2) \, e^{-i\omega_i t_2}
\end{eqnarray}

Substituting the $g$ functions of Eq.~(\ref{g-2}) into Eq.~(\ref{psi-2}) and applying
the $\delta$-functions in the state to reduce the variables of the integral in 
Eq.~(\ref{psi-2}) from 4 to 2, we obtain
\begin{eqnarray}\label{calc}
&&\Psi(\vec{\rho}_1,\tau_1;\vec{\rho}_2,\tau_2)  \nonumber \\
&\propto&\int d\vec{\rho}_{o} \,  A(\vec{\rho}_{o}) \; 
e^{i \frac{\omega_p}{2 c s_o} |\vec{\rho}_{o}|^2} 
\, \int d\vec{\rho'}_{o}  \; A(\vec{\rho'}_{o}) \; e^{i \frac{\omega_p}{2 c s_o} |\vec{\rho'}_{o}|^2} 
\nonumber \\
&\times& \int d\vec{\rho}_l \, e^{-i \frac{\omega_p}{2
c} (\frac{\vec{\rho}_{o}}{s_o} + \frac{\vec{\rho}_1}{s_{i}}) \cdot
\vec{\rho}_l}
\int  d\vec{\rho'}_l \; e^{-i \frac{\omega_p}{2 c} (\frac{\vec{\rho'}_{o}}{s_o} + 
\frac{\vec{\rho}_2}{s_{i}}) \cdot \vec{\rho'}_l} \nonumber \\
&\times& \Big{\{} \int d \nu \: e^{i \nu [(\tau_1 - \tau_2)+(\frac{\vec{\rho}_{o}}{c s_o} + 
\frac{\vec{\rho}_1}{c s_{i}}) \cdot \vec{\rho}_l - (\frac{\vec{\rho'}_{o}}{c s_o} + 
\frac{\vec{\rho}_2}{c s_{i}}) \cdot \vec{\rho'}_l]} \Big{\}} \nonumber \\
&\times& e^{-i \frac{\nu}{2 c s_o} |\vec{\rho}_{o}|^2}
e^{i \frac{\nu}{2 c s_o} |\vec{\rho'}_{o}|^2}  \Big{\{} \int d\vec{\kappa}_s \,
e^{i \vec{\kappa}_s \cdot [\vec{\rho}_{o} - \vec{\rho'}_{o}]} \Big{\}}
\end{eqnarray}
where $\tau_j = t_j - z_{j} /c $, $\nu$ is defined from $\omega_s = \omega_{p}/2 + \nu$ and 
$\omega_s = \omega_{p}/2 - \nu$ following
$\omega_s + \omega_i=\omega_p$.

Now we consider a two-photon film (two-photon absorber), or equivalently scan 
$D_1$ and $D_2$ together, on the image plane to achieve
$\vec{\rho}_{1} = \vec{\rho}_{2}$,  $z_1 = z_2$ and examine 
the two integrals in the two curly brackets.   It is easy to see that the integral of 
$d\vec{\kappa}_s$ yields $\delta(\vec{\rho}_{o} - \vec{\rho'}_{o})$, 
which is consistent with Eq.~(\ref{Wavefunction}).   The integral of $d \nu$ gives a similar
$\delta$-function in the form of $\delta[(\frac{\vec{\rho}_{o}}{c s_o} + 
\frac{\vec{\rho}}{c s_{i}}) (\vec{\rho}_{l} - \vec{\rho'}_{l})]$ while taking 
$\vec{\rho}_{o} = \vec{\rho'}_{o}$, 
$\vec{\rho}_{1} = \vec{\rho}_{2} = \vec{\rho}$, and $\tau_1 = \tau_2$.  These results
indicate that the propagation-diffraction of the signal photon and the idler photon are not 
independent.  The ``two-photon diffraction" couples the 
two integrals in $\vec{\rho}_{o}$ and $\vec{\rho'}_{o}$
as well as the two integrals in $\vec{\rho}_{l}$ and $\vec{\rho'}_{l}$ 
and gives the $G^{(2)}$ function
\begin{eqnarray}\label{eq_image2}
G^{(2)}(\vec{\rho}, \vec{\rho}) 
\propto \Big{|}\int_{obj} d\vec{\rho}_{o} \;
A^2(\vec{\rho}_{o}) \, e^{i \frac{\omega_p}{2 c s_o} |\vec{\rho}_{o}|^2} 
\frac{2 J_{1}\Big{(}\frac{R}{s_o}
\frac{\omega_p}{c} \big{|}\vec{\rho}_{o} +
\frac{\vec{\rho}}{m}\big{|}\Big{)}}{\Big{(}\frac{R}{s_o}
\frac{\omega_p}{c} \big{|}\vec{\rho}_{o} +
\frac{\vec{\rho}}{m}\big{|}\Big{)}^{2}} \Big{|}^2
\end{eqnarray}
which indicates that a coherent image magnified by a factor of 
$m=s_i/s_o$ is reproduced on the image plane by joint-detection or by 
two-photon absorption.

In Eq.~(\ref{eq_image2}), the point-spread function is
characterized by the pump wavelength $\lambda_p = \lambda_{s,i}/2$; 
hence, the point-spread function is twice narrower than in
the (first order) classical case. An entangled two-photon state 
thus gives an image in joint-detection
with double spatial resolution when compared to what one would
obtain in classical imaging. Moreover, the spatial resolution of
the two-photon image obtained by SPDC  is further
improved because it is determined by the function $2 J_1(x)/ x^2$, which is much
narrower than the $somb(x)$.  

It is interesting to see that, different from the classical case,
the frequency integral over $\Delta \omega_s \sim \infty$ does not
give any blurring problem, but rather enhances the spatial
resolution of the two-photon image.  These characteristics
of entangled states may be useful for certain applications 
\cite{aberration}.

Can we replace entangled states with classical light in the same setup of
Fig.~\ref{fig:lithography-1} and still have a twice narrower
point-spread function in two-photon joint-detection?

We will first quickly examine the case of coherent radiation. By
definition, coherent light is characterized by the relation:
$G^{(2)}=|G^{(1)}|^2$ \cite{Glauber}. The two-photon image
produced by laser light is readily obtained from Eq.~(\ref{i-1}):
\begin{eqnarray}\label{laser}
G^{(2)}(\vec{\rho}_1,\vec{\rho}_2) &\propto& 
\Big{|} \int_{obj} d\vec{\rho}_{o} \,
A(\vec{\rho}_{o}) \, e^{i \frac{\omega}{2 c s_o} |\vec{\rho}_{o}|^2} 
somb[\frac{R}{s_o} \, \frac{\omega}{c}
|\vec{\rho}_{o} + \frac{\vec{\rho}_{1}}{m}|] \Big{|}^2 \nonumber \\
 &\times& \Big{|} \int_{obj}
d\vec{\rho}_{o} \, A(\vec{\rho}_{o}) \, e^{i \frac{\omega}{2 c s_o} |\vec{\rho}_{o}|^2} 
somb[\frac{R}{s_o} \, \frac{\omega}{c} |\vec{\rho}_{o} + \frac{\vec{\rho}_{2}}{m}|]
\Big{|}^2
\end{eqnarray}
which is simply a product of two independent first-order coherent images
(see Eq. (\ref{i-1})).

Now we analyze chaotic light.  
The standard form of the $G^{(2)}$-function of
chaotic light is \cite{hbt}:
\begin{equation}\label{G2-3}
G^{(2)}(\vec{\rho}_{1}, \vec{\rho}_{2}) =
G^{(1)}_{11}(\vec{\rho}_{1}) \, G^{(1)}_{22}(\vec{\rho}_{2})+
|G^{(1)}_{12}(\vec{\rho}_{1}, \vec{\rho}_{2})|^{2},
\end{equation}
where $G^{(1)}_{11}(\vec{\rho}_{1})$ and
$G^{(1)}_{22}(\vec{\rho}_{2})$ are the mean intensity
distributions on the image plane. 
It is straightforward to find that the first term of Eq.~(\ref{G2-3}), i.e.,
$G^{(1)}_{11}(\vec{\rho}_{1}) \, G^{(1)}_{22}(\vec{\rho}_{2})$, is
a simple product of two independent incoherent classical
(first-order) images.
The interesting part of Eq.~(\ref{G2-3}) is in the second term where
\begin{eqnarray}\label{G12}
G^{(1)}_{12}(\vec{\rho}_1,z_1, t_1;\vec{\rho}_2, z_2, t_2)  
\propto \int d \vec{\kappa} \, 
g^{*}(\vec{\kappa}, \omega; \vec{\rho}_1, z_1) \, e^{i\omega t_1}  \,
g(\vec{\kappa}, \omega; \vec{\rho}_2, z_2) \, e^{-i\omega t_2}.
\end{eqnarray}
Substituting the $g^{*}$ and $g$ functions into Eq.~(\ref{G12}) and considering
$D_1$ and $D_2$ on the image plane, we obtain
\begin{eqnarray}\label{calc-chaotic}
G^{(1)}_{12}(\vec{\rho}_1, \vec{\rho}_2) &\propto& \int_{obj} d\vec{\rho}_{o} \,  
A^{*}(\vec{\rho}_{o}) \; 
e^{-i \frac{\omega}{2 c s_o} |\vec{\rho}_{o}|^2} 
 \int_{obj} d\vec{\rho'}_{o}  \; A(\vec{\rho'}_{o}) \; 
e^{i \frac{\omega}{2 c s_o} |\vec{\rho'}_{o}|^2} 
\nonumber \\
&\times& \int_{lens} d\vec{\rho}_l \, e^{i \frac{\omega}{c} 
(\frac{\vec{\rho}_{o}}{s_o} + \frac{\vec{\rho}_1}{s_{i}}) \cdot
\vec{\rho}_l}
\int_{lens} d\vec{\rho'}_l \; e^{-i \frac{\omega}{ c} (\frac{\vec{\rho'}_{o}}{s_o} 
+ \frac{\vec{\rho}_2}{s_{i}}) \cdot \vec{\rho'}_l} \nonumber \\
&\times&  \Big{\{}  \int d\vec{\kappa} \,
e^{i \vec{\kappa}  \cdot [\vec{\rho'}_{o} - \vec{\rho}_{o}]} \Big{\}}.
\end{eqnarray}  
Similar to the first-order imaging, in Eqs.~(\ref{G12}) and (\ref{calc-chaotic}) we have 
assumed monochromatic light to avoid ``aberration" complications.  Differing from that 
of SPDC, here an integral of $\Delta \omega \sim \infty$ results in a constant 
$G^{(1)}_{12}$, which is consistent with our experimental observation \cite{prl1}.  

Evaluating the integrals in 
Eq.~(\ref{calc-chaotic}), the $G^{(1)}_{12}G^{(1)}_{21}$ term turns out to be 
\begin{eqnarray}\label{thins}
|G^{(1)}_{12}(\vec{\rho}_{1}, \vec{\rho}_{2})|^{2} \propto 
\Big{|}\int_{obj} d\vec{\rho}_{o} \,
|A(\vec{\rho}_{o})|^2 
\, somb\Big{(}\frac{R}{s_o} \frac{\omega}{c}
\Big{|}\vec{\rho}_{o}+\frac{\vec{\rho}_1}{m} \Big{|} \Big{)}
somb\Big{(}\frac{R}{s_o} \frac{\omega}{c}
\Big{|}\vec{\rho}_{o}+\frac{\vec{\rho}_2}{m} \Big{|} \Big{)} \Big{|}^{2}.
\end{eqnarray}
Unlike the $G^{(1)}_{11}G^{(1)}_{22}$ term, here, we do not have two
independent images.  This is the reason that chaotic light has been
successfully used to mimic certain nonlocal features of entangled
systems \cite{prl2}. Chaotic light,  however, is not a viable
alternative to entangled sources for quantum lithography. In fact,
when $D_1$ and $D_2$ are scanned together, or a two-photon
sensitive material is employed for recording the photon pair,
i.e., $\vec{\rho}_1=\vec{\rho}_2=\vec{\rho}$, Eq.~\ref{thins}
becomes:
\begin{eqnarray}\label{thres}
|G^{(1)}_{12}(\vec{\rho}, \vec{\rho})|^{2} 
\propto
\big{|}\int_{obj} d\vec{\rho}_{o} \; |A(\vec{\rho}_{o})|^2 \;
|somb[\frac{R}{s_o} \frac{\omega}{c} \big{|}\vec{\rho}_{o} +
\frac{\vec{\rho}}{m}\big{|}]|^{2}\Big{|}^2
\end{eqnarray}
which is the same result as given by the $G^{(1)}_{11}G^{(1)}_{22}$
term. Namely, it is simply the product of two independent incoherent
first-order classical images. This conclusion relates quite nicely
to the findings of \cite{Europhys}: in that experiment, a doubly
modulated interference-diffraction pattern was observed by
scanning the detectors in opposite directions on the Fourier
transform plane. However, no second-order interference pattern
can be observed if the detectors are scanned together.

\section{Ghost Imaging}

\hspace{6.5mm}Pittman \emph{et al} demonstrated a surprising two-photon imaging experiment 
in 1995 \cite{GhostImage}.  The experiment was immediately named ``ghost imaging" 
due to its ``nonlocal" feature.   The important physics demonstrated in the experiment, 
nevertheless, may not be the so called ``ghost."  Indeed, the original purpose of 
the experiment was to study the EPR correlation in position and in momentum and 
to test the EPR inequality \cite{IEEE-03,disug,Milena} for the entangled signal-idler 
photon pair of SPDC.   

\begin{figure}[hbt]
    \centering
    \includegraphics[width=88mm]{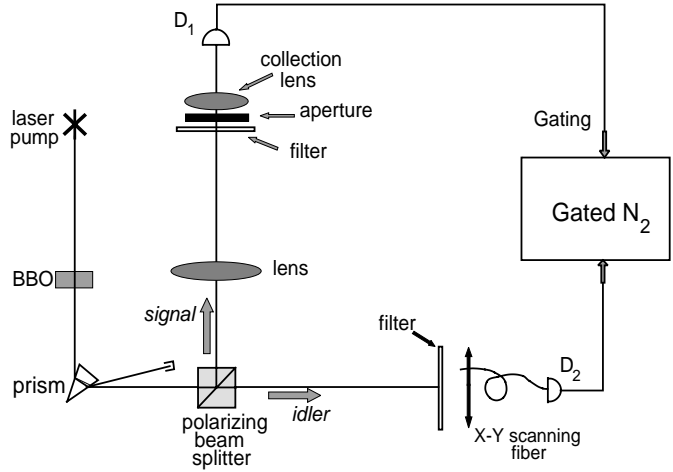}
    {\caption{Schematic set-up of the ``ghost'' image experiment.}
    \label{Imageset}}
\end{figure}  
The schematic setup of the  ``ghost" imaging experiment is shown in 
Fig. \ref{Imageset}. A CW laser is used to pump a nonlinear
crystal, which is cut for degenerate type-II phase matching to produce a pair 
of orthogonally polarized signal (e-ray of the crystal) and idler (o-ray of the crystal) 
photon. The pair emerges from the
crystal collinear, with $\omega _{s}\cong \omega _{i}\cong \omega
_{p}/2$. The pump is then separated from the signal-idler
pair by a dispersion prism, and the remaining
signal and idler beams are sent in different directions by a polarization
beam splitting Thompson prism. The signal beam passes through a convex lens
with a $400mm$ focal length and illuminates a chosen aperture (mask). As an
example, one of the demonstrations used letters ``UMBC'' for the object mask. 
Behind the aperture is the ``bucket'' detector package $D_{1}$, which consists of a 
short focal length collection lens in whose focal spot is an avalanche photodiode.  
$D_{1}$ is mounted in a fixed position during the experiment.  The idler beam is met by 
detector package $D_{2} $, which consists of an optical fiber whose output is mated 
with another avalanche photodiode. The input tip of the fiber is scanned in the transverse 
plane by two step motors. The output pulses of each detector, which are operating in 
photon counting mode, are sent to a coincidence counting circuit for the 
signal-idler joint-detection.  
\begin{figure}[hbt]
    \centering
    \includegraphics[width=80mm]{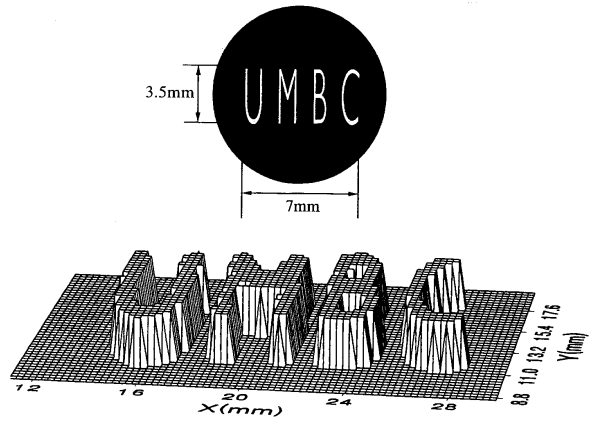}
     \parbox{14cm}{{\caption{(a) A reproduction of the
actual aperature ``UMBC'' placed in the signal beam. (b) The image
of ``UMBC'': coincidence counts as a function of the fiber tip's
transverse coordinates in the image plane. The step size is 0.25mm. The image
shown is a ``slice'' at the half maximum value.}}
    \label{UMBC}}
\end{figure}

By recording the coincidence counts as a function of the fiber tip's
transverse plane coordinates, the image of the chosen aperture (for
example, ``UMBC'') is observed, as reported in Fig. \ref{UMBC}. It is interesting to
note that while the size of the ``UMBC'' aperture inserted in the signal beam is only about 
$3.5mm\times 7mm$, the observed image measures $7mm\times14mm$.  The image is 
therefore magnified by a factor of 2.  The observation also confirms that the focal length 
of the imaging lens $f$, the aperture's optical distance from the lens $S_{o}$, and the 
image's optical distance from the lens $S_{i}$ (which is from the imaging lens going 
backward along the signal photon path to the two-photon source of SPDC crystal then 
going forward along the path of idler photon to the image), satisfy the Gaussian thin lens 
equation.  In this experiment, $S_{o}$ was chosen to be $S_{o}=600mm$, and the twice 
magnified clear image was found when the fiber tip was on the plane of 
$S_{i}=1200mm$.  When $D_2$ was scanned on transverse planes not determined 
by the Gaussian thin lens equation the images blurred out.

The measurement of the signal and the idler subsystem themselves 
are very different.  The single 
photon counting rate of $D_{2}$ was recorded during the scanning of the image and was 
found fairly constant in the entire region of the image.   This means that the transverse 
coordinate uncertainty of either signal or idler is considerably large compared to that of 
the transverse correlation of the entangled signal-idler photon pair: 
 $\Delta x_{1}$ ($\Delta y_{1}$) and $\Delta x_{2}$ ($\Delta y_{2}$) are much 
greater than $\Delta (x_{1} - x_{2})$ ($\Delta (y_{1} - y_{2})$).

\begin{figure}[hbt]
    \centering
    \includegraphics[width=80mm]{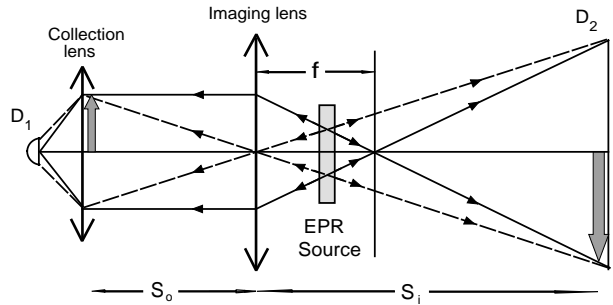}
     \parbox{14cm}{\caption{An unfolded setup of the ghost imaging experiment, 
    which is helpful for understanding the physics.  Since the two-photon ``light" propagates along 
    ``straight-lines", it is not difficult to find that any geometrical light point on the object plane 
    corresponds to an unique geometrical light point on the image plane.  Thus, a ghost image 
    of the object is made nonlocally in the image plane.  Although the placement of the lens, the 
    object, and detector $D_{2}$ obeys the Gaussian thin lens equation, it is important to 
    remember that the geometric rays in the figure actually represent the two-photon amplitudes 
    of an entangled photon pair.   The point 
    to point correspondence is the result of the superposition of these two-photon amplitudes.} 
    \label{fig:imaging-unfold}}
\end{figure}

The EPR $\delta$-functions, $\delta(\vec{\rho}_s - \vec{\rho}_i)$ and
$\delta(\vec{\kappa}_s + \vec{\kappa}_i)$ in transverse dimension, are the key to 
understand this interesting phenomenon.  In degenerate SPDC, although the 
signal-idler photon pair has equal
probability to be emitted from any points on the output surface of the nonlinear crystal, the 
transverse position $\delta$-function indicates that if one of them is observed at one position,
the other one must be found at the same position.  In other words, the pair is always emitted
from the same point on the output plane of the two-photon source.  Simultaneously, 
the transverse momentum $\delta$-function
defines the angular correlation of the signal-idler pair.  The transverse momenta of a 
signal-idler amplitude are always equal but pointed in opposite directions,
$\vec{\kappa}_s =-\vec{\kappa}_i$, which means that 
the two-photon amplitudes are always existing at roughly equal 
yet opposite angles relative to the pump.  This then allows for a simple
explanation of the experiment in terms of ``usual'' geometrical optics in
the following manner: we envision the nonlinear crystal as a ``hinge point'' and
``unfold'' the schematic of Fig. \ref{Imageset} into the Klyshko picture of 
Fig. \ref{fig:imaging-unfold}.  The signal-idler two-photon amplitudes can then be 
represented by straight lines (but keep in mind the different propagation directions) 
and therefore, the image is well produced in coincidences when the aperture, lens, 
and fiber tip are located according to the Gaussian thin lens equation of 
Eq.~(\ref{Lens-Eq}).  The image is exactly the same as one
would observe on a screen placed at the fiber tip if detector $D_{1}$ were
replaced by a point-like light source and the nonlinear crystal by a reflecting
mirror. 

Following a similar analysis in geometric optics,  it is not difficult to find that any 
geometrical ``light spot" on the object plane, which is the intersection point of all 
possible two-photon amplitudes coming from the two-photon light source, 
corresponds to an unique geometrical ``light spot" on the image 
plane, which is another intersection point of all the possible two-photon amplitudes.  
This point to point correspondence made the ``ghost" image of the subject-aperture possible.  
Despite the completely different physics from classical geometrical optics, the remarkable 
feature is that the relationship between the focal length of the lens $f$, the aperture's 
optical distance from the lens $S_{o}$, and the image's optical distance from the lens 
$S_{i}$, satisfy the Gaussian thin lens equation of Eq.~(\ref{Lens-Eq}).
Although the placement of the lens, the object, and the detector $D_{2}$ obeys the Gaussian 
thin lens equation, it is important to remember that the geometric rays in the figure actually 
represent the two-photon amplitudes of a signal-idler photon pair and the point to point 
correspondence is the result of the superposition of these two-photon amplitudes.  
The ``ghost" image is a realization of the 1935 EPR {\em gedankenexperiment}.  

Now we calculate $G^{(2)}(\vec{\rho}_o, \vec{\rho}_i)$ for the ``ghost" imaging 
experiment, where $\vec{\rho}_o$ and $ \vec{\rho}_i$ are the transverse coordinates 
on the object plane and the image plane.  We will show that there exists a $\delta$-function 
like point-to-point correlation between the object plane and the image plane, i.e., if one 
measures the signal photon at a position of $\vec{\rho}_o$ on the object plane the idler 
photon can be found only at a certain unique position of $\vec{\rho}_i$ on the image plane 
satisfying $\delta(\vec{\rho}_o - \vec{\rho}_i / m)$, where $m=-(s_i/s_o)$ is the image-object 
magnification factor.  After approving the $\delta$-function correlation, we show how the object 
function of $A(\vec{\rho}_o)$ is transferred to the image plane as a magnified image 
$A(\vec{\rho}_i/m)$.  Before starting the calculation, it is worth to emphasize again 
that the ``straight lines" in Fig.~\ref{fig:imaging-unfold} schematically represent 
the two-photon amplitudes all belong to a pair of signal-idler photon.  
A ``click-click" joint measurement at ($\mathbf{r}_{1}, t_{1}$), which is on the object
plane, and ($\mathbf{r}_{2}, t_{2}$), which is on the image plane, in the form of 
EPR $\delta$-function, is the result of the coherent superposition of all these
two-photon amplitudes.  

\begin{figure}[hbt]
    \centering
    \includegraphics[width=90mm]{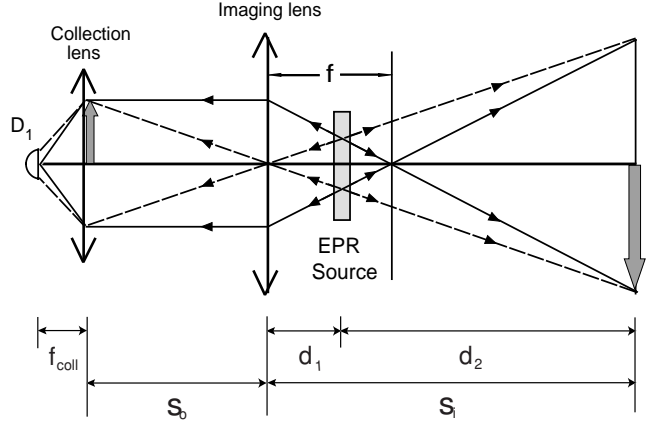}
     \parbox{14cm}{\caption{In arm-$1$, the signal propagates freely over a distance $d_1$ 
    from the output plane of the source to the imaging lens, then passes an object aperture 
    at distance $s_o$, and then is focused onto photon counting detector $D_1$ by a 
    collection lens.  In arm-$2$, the idler propagates freely over a distance $d_2$ from the 
    output plane of the source to a point-like photon counting detector $D_2$.} 
    \label{fig:imaging-unfold-2}}
\end{figure}

We follow the unfolded experimental setup shown in Fig.~\ref{fig:imaging-unfold-2} to 
establish the Green's functions $g(\vec{\kappa}_s, \omega_s, \vec{\rho}_o, z_o)$ and 
$g(\vec{\kappa}_i, \omega_i, \vec{\rho}_2, z_2)$.  In arm-$1$, the signal propagates freely 
over a distance $d_1$ from the output plane of the source to the imaging lens, then 
passes an object aperture at distance $s_o$, and then is focused onto photon 
counting detector $D_1$ by a collection lens.  We will evaluate 
$g(\vec{\kappa}_s, \omega_s, \vec{\rho}_o, z_o)$ by propagating the field from 
the output plane of the two-photon source to the object plane.
In arm-$2$, the idler propagates freely over a distance $d_2$ 
from the output plane of the two-photon source to a point-like detector $D_2$.     
$g(\vec{\kappa}_i, \omega_i, \vec{\rho}_2, z_2)$ is thus a free propagator.

\vspace{3mm}
\hspace*{-4mm}(I) Arm-$1$ (source to object):
\vspace{1mm}

The optical transfer function or Green's function in arm-$1$, which propagates the field 
from the source plane to the object plane, is given by:
\begin{eqnarray}\label{Arm-1}
&& g(\vec{\kappa}_s, \omega_s; \vec{\rho}_o, z_o = d_{1}+s_{o}) \nonumber \\
&=& e^{i \frac{\omega_s}{c} z_o} 
\int_{lens} d\vec{\rho}_l \, \int_{source} d\vec{\rho}_s 
\, \Big{\{} \, \frac{-i \omega_s}{2 \pi c d_1} e^{i \vec{\kappa_s} \cdot \vec{\rho}_s}  
\, e^{i \frac{\omega_s}{2 c d_1} |\, \vec{\rho}_s-\vec{\rho}_l\, |^2 } \Big{\}} \nonumber \\ 
& & \times \,\, e^{-i \frac{\omega}{2 c f} \, | \vec{\rho}_l |^2} \, 
\Big{\{} \, \frac{-i \omega_s}{2 \pi c s_{o}} \,
e^{i \frac{\omega_s}{2 c s_o} |\,\vec{\rho}_l-\vec{\rho}_o\,|^2} \Big{\}},
\end{eqnarray}
where $\vec{\rho}_s$ and $\vec{\rho}_l$ are the transverse vectors defined, respectively, 
on the output plane of the source and on the plane of the imaging lens.  
The terms in the first and second curly brackets in Eq.~(\ref{Arm-1}) describe free space 
propagation from the output plane of the source to the imaging lens and from the imaging 
lens to the object plane, respectively.  Again, 
$ e^{i \frac{\omega_s}{2 c d_1} |\, \vec{\rho}_s-\vec{\rho}_l\, |^2 } $ and 
$e^{i \frac{\omega_s}{2 c s_o} |\,\vec{\rho}_l-\vec{\rho}_o\,|^2} $  are
the Fresnel phases we have defined in Appendix A.  Here, we treat the imaging lens 
as a thin-lens. Thus, the transformation function of the imaging lens is approximated as 
a Gaussian $l(| \vec{\rho}_l |, f) \cong e^{-i \frac{\omega}{2 c f} \, | \vec{\rho}_l |^2}$.

\vspace{3mm}
\hspace{-4mm}(II) Arm-$2$ (from source to image):
\vspace{2mm}

In arm-$2$, the idler propagates freely from the source to the plane of $D_2$, which
is also the plane of the image. The Green's function is thus:
\begin{eqnarray}\label{Arm-2} 
g(\vec{\kappa}_i, \omega_i; \vec{\rho}_2, z_2=d_2) 
= \frac{-i \omega_i}{2 \pi c d_2} \, e^{i \frac{\omega_i}{c} d_2} 
\int_{source} d\vec{\rho'_s}  \,
e^{i \frac{\omega_i}{2 c d_2} |\, \vec{\rho'_s}-\vec{\rho}_2\, |^2} \, 
e^{i \vec{\kappa}_i \cdot \vec{\rho'_s} }
\end{eqnarray}
where $\vec{\rho'_s}$ and $\vec{\rho}_2$ are the transverse vectors defined, respectively, 
on the output plane of the source, and on the plane of the photodetector $D_2$.  

\vspace{3mm}
\hspace{-4mm}(III) $\Psi(\vec{\rho}_o, \vec{\rho}_i)$ and 
$G^{(2)}(\vec{\rho}_o, \vec{\rho}_i)$ (object plane - image plane):
\vspace{1mm}

To simplify the calculation and to focus on the transverse correlation, 
in the following calculation we assume degenerate 
($\omega_s = \omega_i = \omega$) and collinear SPDC.  The transverse 
two-photon effective wavefunction $\Psi(\vec{\rho}_o, \vec{\rho}_2)$ 
is then evaluated by substituting the Green's functions 
$g(\vec{\kappa}_s, \omega; \vec{\rho}_o, z_o) $ and 
$g(\vec{\kappa}_i, \omega; \vec{\rho}_2, z_2)$
into the expression given in Eq.~(\ref{psi-2}):
\begin{eqnarray}\label{biphoton_x}
\Psi(\vec{\rho}_o,\vec{\rho}_2)
&\propto&  \int d\vec{\kappa}_s \, d\vec{\kappa}_i \, 
\delta(\vec{\kappa}_s +  \vec{\kappa}_i) \, g(\vec{\kappa}_s, \omega; \vec{\rho}_o, z_o)  \, 
g(\vec{\kappa}_i, \omega; \vec{\rho}_2, z_2) \nonumber \\
&\propto& e^{i \frac{\omega}{c} (s_o+s_i)}  \int d\vec{\kappa}_s \, d\vec{\kappa}_i \, 
\delta(\vec{\kappa}_s +  \vec{\kappa}_i) \nonumber \\
& \times&  \int_{lens} d\vec{\rho}_l \, \int_{source} d\vec{\rho}_s \,
 \, e^{i \vec{\kappa_s} \cdot \vec{\rho}_s}  
e^{i \frac{\omega}{2 c d_1} |\, \vec{\rho}_s-\vec{\rho}_l\, |^2 }  \nonumber \\ 
& & \hspace*{27.5mm} \times \,\, e^{-i \frac{\omega}{2 c f} \, | \vec{\rho}_l |^2} \,  \, 
e^{i \frac{\omega_s}{2 c s_o} |\,\vec{\rho}_l-\vec{\rho}_o\,|^2}  \nonumber\\
&\times& \int_{source} d\vec{\rho'_s} \,\,
e^{i \vec{\kappa}_i \cdot \vec{\rho'_s} } \,
e^{i \frac{\omega_i}{2 c d_2} |\, \vec{\rho'_s}-\vec{\rho}_2\, |^2}
\end{eqnarray}
where we have ignored all the proportional constants.
Completing the double integral of $d\vec{\kappa}_s$ and $d\vec{\kappa}_s$
\begin{eqnarray}\label{delta-source}
\int d\vec{\kappa}_s \, d\vec{\kappa}_i \, \delta(\vec{\kappa}_s +  \vec{\kappa}_i)\,
e^{i \vec{\kappa_s} \cdot \vec{\rho}_s} \, e^{i \vec{\kappa}_i \cdot \vec{\rho'_s} } 
\sim \, \delta(\vec{\rho}_s - \vec{\rho'_s}),
\end{eqnarray}  
Eq.~(\ref{biphoton_x}) becomes:
\begin{eqnarray}\label{biphoton_y}
&& \Psi(\vec{\rho}_o,\vec{\rho}_2) \\ \nonumber 
&\propto& e^{i \frac{\omega}{c} (s_0+s_i)}
 \int_{lens} d\vec{\rho}_l \int_{source} d\vec{\rho}_s \,
e^{i \frac{\omega}{2 c d_2} |\, \vec{\rho}_2-\vec{\rho}_s\, |^2 } \,
e^{i \frac{\omega}{2 c d_1} |\, \vec{\rho}_s-\vec{\rho}_l\, |^2 } 
\, e^{-i \frac{\omega}{2 c f} \, | \vec{\rho}_l |^2}  \,
e^{i  \frac{\omega}{2 c s_o} |\,\vec{\rho}_l-\vec{\rho}_o\,|^2}.
\end{eqnarray}
We then complete the integral on $d\vec{\rho}_s$,
\begin{eqnarray}\label{biphoton_z}
\Psi(\vec{\rho}_o,\vec{\rho}_2) 
\propto e^{i \frac{\omega}{c} (s_0+s_i)} 
\int_{lens} d\vec{\rho}_l \, 
e^{i \frac{\omega}{2 c s_i} |\, \vec{\rho}_2-\vec{\rho}_l\, |^2} \,
 e^{-i \frac{\omega}{2 c f} \, | \vec{\rho}_l |^2}
e^{i \frac{\omega}{2 c s_o} |\,\vec{\rho}_l-\vec{\rho}_o\,|^2},
\end{eqnarray}
where we have replaced $d_1+d_2$ with $s_i$ (as depicted in Fig.~\ref{fig:imaging-unfold-2}).
Although the signal and idler propagate to different directions along two optical arms, 
Interestingly, the Green function in Eq.~(\ref{biphoton_z}) is equivalent to that of a 
classical imaging setup, if we imagine the fields start propagating from a point $\vec{\rho}_o$ 
on the object plane to the lens and then stop at point $\vec{\rho}_2$ 
on the imaging plane ($\vec{\rho}_2 = \vec{\rho}_i$).  The mathematics is consistent with 
our previous qualitative analysis of the experiment.  

The integral on $d\vec{\rho}_l$ yields a point-to-point relationship between
the object plane and the image plane that is defined by the Gaussian thin-lens 
equation:
\begin{eqnarray}\label{biphoton_zz}
 \int_{lens} d\vec{\rho}_l \, e^{i \frac{\omega}{2 c}
[\frac{1}{s_o} + \frac{1}{s_i} - \frac{1}{f}] |\, \vec{\rho}_l|^2}  \,
e^{-i\frac{\omega}{c} (\frac{\vec{\rho}_o}{s_o} + \frac{\vec{\rho}_i}{s_i})\cdot \vec{\rho}_l} 
\sim \delta(\vec{\rho}_o + \vec{\rho}_i / m),
\end{eqnarray}
where we have replaced $\vec{\rho}_2$ with $\vec{\rho}_i$.  In Eq.~(\ref{biphoton_zz}),
the integral is approximated to infinity and the Gaussian thin-lens 
equation of Eq~(\ref{Lens-Eq}) is applied.  We have also defined $m=s_i/s_o$
as the magnification factor of the imaging system.  
The function $\delta(\vec{\rho}_o + \vec{\rho}_i / m)$ indicates that a point of 
$\vec{\rho}_o$ on the object plane corresponds to a unique point of 
$\vec{\rho}_i$ on the image plane.  The two vectors pointed to opposite directions and 
the magnitudes of the two vectors hold a ratio of $m=|\vec{\rho}_i|/|\vec{\rho}_o|$.  

If the finite size of the imaging lens has to be taken into 
account (finite radius $R$), the integral yields a point-spread function of 
$somb(x)$:
\begin{eqnarray}\label{somb}
\int_{lens} d\vec{\rho}_l \,
e^{-i\frac{\omega}{c} (\frac{\vec{\rho}_o}{s_o} + \frac{\vec{\rho}_i}{s_i})\cdot 
\vec{\rho}_l} \propto somb\Big{(}\frac{R}{s_o}\, \frac{\omega}{c} 
[\vec{\rho}_{o} + \frac{\vec{\rho}_{i}}{m}] \Big{)}
\end{eqnarray}
where, again, $somb(x) = 2J_1(x)/x$, $J_1(x)$ is the first-order Bessel function.  
The point-spread function turns 
the point-to-point correspondence between the object plane and the image plane into 
a point-to-``spot" relationship and thus limits the spatial resolution.    

Therefore, by imposing the condition of the Gaussian thin-lens equation, the transverse 
two-photon effective wavefunction is approximated as a $\delta$ function 
\begin{eqnarray}\label{biphoton_x_fin}
\Psi(\vec{\rho}_o,\vec{\rho}_i) \sim
\delta(\vec{\rho}_o + \vec{\rho}_i / m),
\end{eqnarray}
which indicates a point to point EPR 
correlation between the object plane and the image plane, i.e., if one 
observes the signal photon at a position of $\vec{\rho}_o$ on the object plane, 
the idler photon can only be found at a certain unique position of $\vec{\rho}_i$ 
on the image plane satisfying $\vec{\rho}_o + \vec{\rho}_i / m =0$ with $m=s_i/s_o$. 

We now include an object-aperture function, a collection lens and a photon counting
detector $D_1$ into the optical transfer function of arm-$1$ as shown in Fig.~\ref{Imageset}.   
The collection-lens-$D_1$ package can be simply treated as a 
``bucket" detector.  The ``bucket" 
detector integrates all $\Psi(\vec{\rho}_o, \vec{\rho}_2)$ that pass 
the object aperture $A(\vec{\rho}_o)$ as a joint photodetection event.  This process 
is equivalent to the following convolution:
\begin{eqnarray}\label{biphoton_final-2}
R_{1, 2} \propto  \int_{object} d\vec{\rho}_o \, \big{|}  A(\vec{\rho}_o)  \big{|}^2 \,
\big{|}  \Psi(\vec{\rho}_o, \vec{\rho}_i)  \big{|}^2  
\simeq \big{|} A(\vec{\rho}_i / m) \big{|}^{2}
\end{eqnarray}
where, again, $D_2$ is scanning in the image plane, $\vec{\rho}_2 = \vec{\rho}_i$. 

As we have discussed earlier, the position-position EPR correlation is the result 
of the coherent superposition of two-photon amplitudes.
\begin{eqnarray}\label{Coherent-Img}
G^{(2)}(\vec{\rho}_o,\vec{\rho}_i) 
= \Big{|}  \int d\vec{\kappa}_s \, d\vec{\kappa}_i \, 
\delta(\vec{\kappa}_s +  \vec{\kappa}_i) \, g(\vec{\kappa}_s, \vec{\rho}_o)  \, 
g(\vec{\kappa}_i, \vec{\rho}_2) \Big{|}^2
\end{eqnarray}
In principle, one signal-idler pair contains all the necessary two-photon amplitudes 
that generate the point-to-point correspondence between the object and the ghost image plane. 
To emphasize this concept, we name this kind of ghost image as 
\textit{two-photon coherent} image.

\begin{figure}[htb]
 \centering
    \includegraphics[width=85mm]{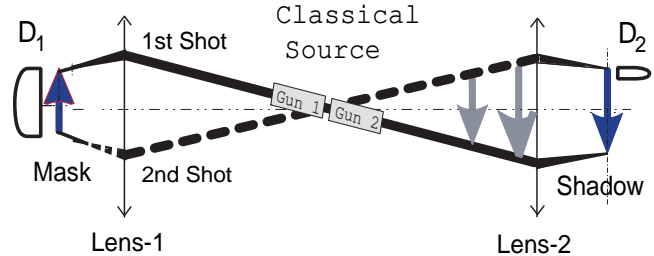}
    \parbox{14cm}{\caption{The projection shadow of an object mask is reproduced in coincidences 
   between $D_1$ and $D_2$ by the use of two co-rotating correlated laser beams.  
   The object-shadow correspondence is an accumulation of the 
   ``momentum-momentum" correlation between each pair of ``shots".} 
   \label{fig:boyd}}
\end{figure}

\vspace{3mm}
We emphasize again, one should not confuse a classical
``momentum-momentum" correlation with the EPR correlation in position 
and in momentum.  Figure \ref{fig:boyd} is a schematic picture of the experiment of 
Bennink \emph{et al}. \cite{boyd}, which distinguishes a trivial classical 
momentum-momentum correlation from EPR.  

\vspace{2mm}

\section{Ghost Imaging of Chaotic Light}

\hspace{6.5mm}Unlike first-order correlation, which is considered as a 
coherent effect of the electromagnetic field, the second-order correlation
of radiation is usually considered as the statistical correlation of 
intensity fluctuations. The first set of second-order correlation  
of thermal light was demonstrated in 1956 by Hanbury Brown
and Twiss (HBT) with two different type of correlations: temporal
and spatial \cite{hbt}.   The HBT experiment created quite a
surprise in the physics community with an enduring debate about
the classical or quantum nature of the phenomenon. It has been
popular to consider that the HBT experiment measures the classical
statistical correlation of the intensity fluctuations of the
radiation \cite{Speckle}:
\begin{eqnarray}\label{Intensity-1}
\langle \, \Delta I_{1} \Delta I_{2} \, \rangle
=  \langle (I_{1} - \bar{I}_{1}) (I_{2} - \bar{I}_{2}) \rangle
= \langle \, I_{1} I_{2} \, \rangle
- \bar{I}_{1} \bar{I}_{2}
\end{eqnarray}
where $\bar{I}_1$ and $\bar{I}_2$ are the mean intensities of the radiation 
measured by photodetectors $D_1$ and $D_2$, respectively.

\begin{figure}[htb]
 \centering
    \includegraphics[width=80mm]{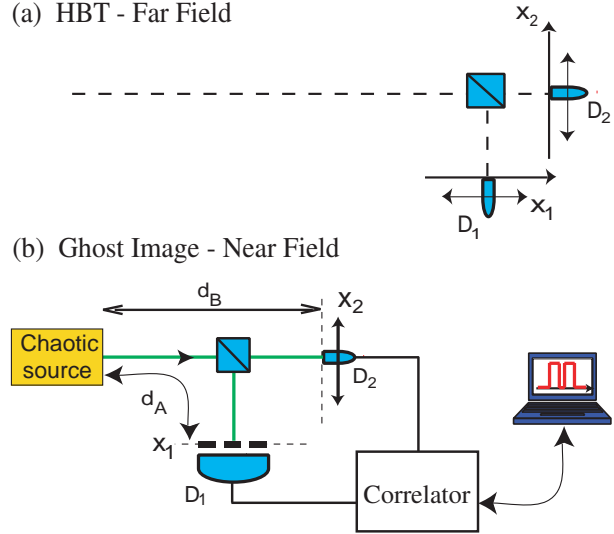}
    \parbox{14cm}{\caption{(a) Hanbury Brown and Twiss configuration. (b) lens-less Ghost
imaging configuration.}
    \label{fig:HBT-Fig1}}
\end{figure}

Figure~\ref{fig:HBT-Fig1} (a) is a schematic of the historical HBT
experiment which measures the second-order transverse spatial
correlation of a monochromatic radiation of wavelength $\lambda$
coming from a distant star with an angular size of $\Delta
\theta$.  Both photodetectors $D_1$ and $D_2$ can be scanned along
the axes $x_1$ and $x_2$, here, we assume 1-D scanning to simplify 
the discussion.  The second-order transverse spatial correlation 
function $\Gamma^{(2)}(x_1, x_2)$ is expected to be
\begin{eqnarray}\label{HBT-1}
\Gamma^{(2)}(x_1, x_2) 
= \langle \, I_{1} \,  I_{2} \, \rangle \sim I_{0}^{2} \,
\Big\{1+sinc^{2}[\,\frac{\pi\Delta\theta(x_1+x_2)}{\lambda}\,]\Big\}
\end{eqnarray}
where we have simplified the problem to $1$-D and assumed
$\bar{I}_{1}=\bar{I}_{2}=I_{0}$. The second term in
Eq.~(\ref{HBT-1}), $I_{0}^{2} \,
sinc^{2}[\pi\Delta\theta(x_1+x_2)/\lambda]$, is interpreted as the
correlation of intensity fluctuations.  This term is useful in
astronomy for angular size measurement of stars.  For short
wavelengths, this function quickly drops from its maximum to
minimum when $x_1+x_2$ goes from zero to a value such that
$\Delta\theta(x_1+x_2)/\lambda=1$. Thus, we effectively have a
``point" to ``point" relationship between the $x_1$ and $x_2$
plane, i.e., for each positive (negative) value of $x_1$ there
exist only one negative (positive) value of $x_2$ that may have
nonzero correlation of intensity fluctuation. In fact,
the planes of $x_1$ and $x_2$ are in the far-field zone
of the finite-size distant star (equivalent to the Fouier transform 
plane).  Therefore,  the measured
quantity is the correlation between the transverse $\mathbf{k}$
vectors of the radiation.  For a narrow function, the non-zero
correlation corresponds to the case of equal transverse
wavevectors: $\vec{\kappa}_1=\vec{\kappa}_2$. This is consistent
with the physics behind the model of classical correlation.  It is
natural to imagine that the radiation coming from the same mode of
the electromagnetic field, passing through the same optical path,
would have identical intensity fluctuations, while radiation
coming from different modes, passing through different optical
paths would not share the same intensity fluctuations.

Quantum models of HBT experiment
derives the same correlation function \cite{fano}\cite{scully}.  However,
if ``classical" works, why quantum?  The classical statistical
interpretation has been widely accepted.  Moreover, the concept of
intensity fluctuation correlation has been even extended to quantum models to
take over the concept of two-photon coherence.  The philosophy of
``photon bunching" is essentially a phenomenological extension to
quantum theory of the statistical correlation on photon number
fluctuations.

In the past twenty years, the massive research on quantum
entanglement has brought new challenges to
the classical statistical correlation interpretation. For example,
replacing the chaotic light with an EPR type two-photon entangled
state in Fig.~\ref{fig:HBT-Fig1} (a), the second-order transverse spatial
correlation function turns out to be
\begin{equation}\label{HBT-2}
\langle \, I_{1} \,  I_{2} \, \rangle \sim I_{0}^{2} \,
sinc^{2}[\,\frac{\pi\Delta\theta(x_1-x_2)}{\lambda}\,].
\end{equation}
Based on the concept of classical statistical correlation of
intensity fluctuation, the mean intensities $\bar{I}_{1}$ and
$\bar{I}_{2}$ must be zero in this case, otherwise
Eq.~(\ref{Intensity-1}) leads to non-physical conclusions.   The
measurements, however, never yield zero mean values of
$\bar{I}_{1}$ and $\bar{I}_{2}$ in any circumstances.

Thus the concept of classical statistical correlation of intensity
fluctuation may not work for entangled two-photon states.
Two-photon correlation experiments with entangled photons have
been explained in terms of the superposition of indistinguishable
alternatives, i.e., two-photon probability amplitudes, that can lead to
a joint-detection event \cite{IEEE-03}. Such alternatives,
however, represent a troubling concept in classical theories,
because this concept has no counterpart in classical electromagnetic 
theory of light and it is nonlocal. If accepted, the nonlocal behavior of
the radiation has been classified as a peculiar property of
non-classical sources.

More interestingly, we further ask ourselves: does the statistical correlation 
of intensity fluctuation always work for chaotic light? We are going to 
analyze a recent ghost imaging experiment of chaotic light by 
Scarcelli \emph{et al.}, 
aimed at answering this question: ``Can two-photon correlation of 
chaotic light be considered as correlation of intensity fluctuations?" \cite{prl2}
We will conclude that the second-order correlation of chaotic light
is a \emph{two-photon interference} phenomenon.

Figure~\ref{fig:HBT-Fig1} (b) illustrates the setup of the experiment. 
Radiation from a chaotic pseudothermal source \cite{martienssen} 
was divided in two optical
paths by a non-polarizing beam splitter. In arm $A$ an object, a
double slit with slit separation $d=1.5 mm$ and slit width $a=0.2
mm$, was placed at a distance $d_{A}=139mm$ and a bucket detector
($D_{1}$) was just behind the object. In arm $B$ a point detector
$D_{2}$ scanned the transverse planes in the neighborhood of a
distance $d_{B}=d_{A}$ from the source.  The correlation was
measured by either photon-counting-coincidence circuit or by standard
HBT correlator.  In the photon counting regime, two Geiger mode avalanche 
photodiodes were employed for single-photon level measurement.  
In HBT scheme, the two photodetectors were Silicon PIN photodiodes.
The bucket detector $D_{1}$ was simulated by using a short focal
length lens ($f=25mm$) to focus the light onto the active area of
the detector while the point detector $D_{2}$ was simulated by a
pinhole. 

\begin{figure}[htb]
 \centering
    \includegraphics[width=70mm]{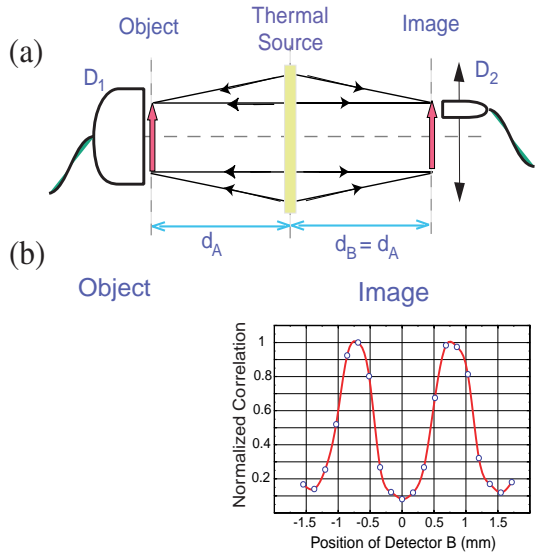}
    \parbox{14cm}{\caption{The results of Lensless imaging. Normalized two-photon 
   correlation vs transverse position of $D_{2}$. The DC constant is
subtracted from the correlation.}
    \label{fig:HBT-Fig2}}
\end{figure}

Figure~\ref{fig:HBT-Fig2} reports the measured two-photon image of the
double-slit. The result shows an equal-size reproduction of the double slit 
when scanning photodetector $D_{2}$ along $x_2$ axis, which is located 
at distance $d_{B}=d_{A}$ from the source.  In Fig.~\ref{fig:HBT-Fig2}, the DC 
background is subtracted from the image.  Here, again, 
we have made a 1-D scanning to simplify the discussion.

Let us first clarify the main differences of this measurement
compared to Hanbury Brown and Twiss types of experiments.
As we described in the beginning of this section, the
measurements of HBT are in the \textit{far-field} zone, which
measure the \textit{momentum-momentum} correlation of the field.
In the reported experiment, instead, the measurements were in the \textit{near
field} zone ($\Delta \theta \sim 10 \lambda/d$) and therefore 
effectively measured the \textit{position-position} correlation
between the object plane and the image plane.  The observation is
a lensless two-photon ghost image.

\begin{figure}[hbt]
 \centering
    \includegraphics[width=85mm]{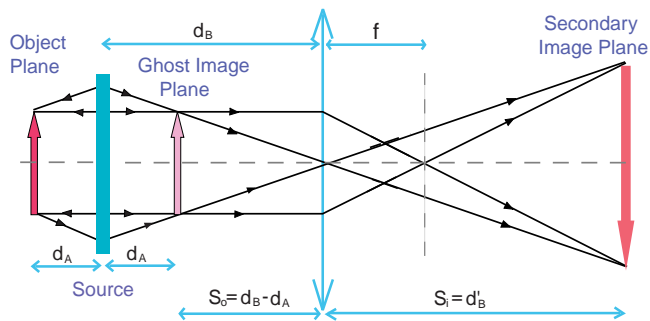}
     \parbox{14cm}{\caption{Schematic experimental setup of a secondary image measurement.}
    \label{fig:Thermal}}
\end{figure}

\begin{figure}[hbt]
 \centering
    \includegraphics[width=67mm]{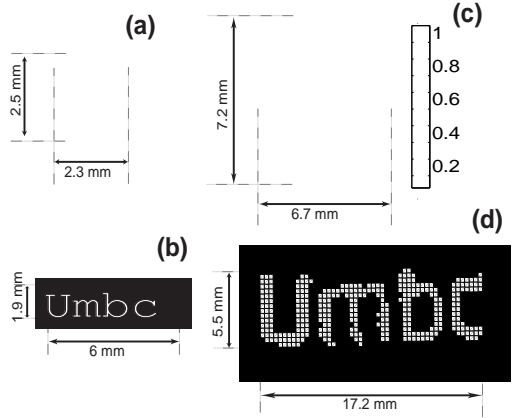}
     \parbox{14cm}{\caption{Results of the secondary Imaging. On the left, (a) and (b) are
the two object used, on the right are shown the obtained images:
(c) is a density plot of the normalized correlation, (d) is a
``slice" of the measured correlation around the half maximum.}
    \label{fig:HBT-Fig3}}
\end{figure}

To confirm that the observed results corresponded to an image, Scarcelli \emph{et al}.
constructed a secondary imaging system, which is illustrated in 
Fig. \ref{fig:Thermal}, by using a $f=85mm$ 
lens to image the ghost image onto a secondary image plane.
The imaging lens was located at a distance of $d_{B}\sim253mm$ from
the source.  A magnified secondary image was observed at
$d'_{B}\sim330 mm$ from the lens, which is given by the Gaussian
thin lens equation, by scanning $D_{2}$ on the transverse plane.
Fig.~\ref{fig:HBT-Fig3} reports the measured magnified secondary image of
the ghost image with the expected magnification
$M=d'_{B}/(d_{B}-d_{A}) \sim2.9$. For this examination  this experiment used two
masks with more complicated structures: one, Fig.~\ref{fig:HBT-Fig3}(a),
with the starting letter of the cities of the authors ($2.3$ x $2.5$ $mm$) and
the other, Fig.~\ref{fig:HBT-Fig3}(b), with the acronym of their institution
($6$ x $1.9$ $mm$). Fig.~\ref{fig:HBT-Fig3}(c) shows the image obtained
with the actual revealed correlation measurements to show the high
contrast of the image, while in Fig.~\ref{fig:HBT-Fig3}(d) is shown a
``slice" of the image around the half maximum of the correlation.

The lens-less ghost imaging setup of Scarcelli \emph{et al}. is a 
straightforward modification of the HBT.  One needs simply 
move the two HBT photodetector from far-field to near-field.
We cannot stop to ask: What has been preventing 
this simple move for 50 years (1956-2006)?  Something must be 
terribly misleading to give us such misled confidence not even to 
try the near-field measurement in half a century. 

\begin{figure}[hbt]
    \centering
     \includegraphics[width=80mm]{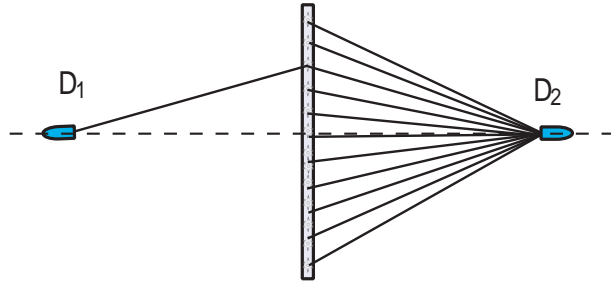}
     \parbox{14cm}{\caption{A simplified model of near-field joint detection of chaotic source.  
    It is easy to see that the chance of receiving radiation from the same sub-source 
    is less then that of receiving from different sub-sources.} 
    \label{fig:multi-mode}}
\end{figure}

The classical model in terms of statistical correlation of intensity fluctuations 
would not work for Scarcelli's experiment. Unlike the HBT experiment 
in which the correlation was measured in the far-field zone.  
Scarcelli's experiment is in near-field.  In the near-field configuration, 
any point on the object plane and on the
image plane are ``hit" by many different $\mathbf{k}$ vectors (modes).  
Fig. \ref{fig:multi-mode} schematically illustrates the situation. Assuming a 
chaotic radiation source of finite dimension consists of a large number of 
$N$ point sub-sources.  By definition of ``chaotic", each point 
sub-sources radiate and fluctuate independently and randomly. Now, we place two 
point photodetectors in near-field as illustrated in Fig. \ref{fig:multi-mode} and
make a similar correlation measurement as that of HBT.  It is easy to see that 
the chance of receiving 
radiation from the same sub-source is less then that of receiving from different 
sub-sources.  The ratio between the two cases is approximately $N/N^{2}$. 
For a large number of $N$, the contributions from ``identical sub-source" is negligible 
and thus $\langle\Delta I_{1} \Delta I_{2}\rangle=0$,
as we know that different sub-sources of chaotic light fluctuate randomly and
independently.  HBT is a special case.  The two photodetectors are placed
in far-field.  A far-field plane is equivalent to the Fourier transform plane.  
On the Fourier transform plane, each point corresponds to a mode of radiation,
and corresponds to a point sub-source.  Thus, the chance for the two 
photodetectors to receive ``identical mode", or receive radiations from 
``identical point sub-sources", is non-negligible.  In this situation, 
$\langle\Delta I_{1} \Delta I_{2}\rangle \neq 0$.  

On the other hand, the quantum model for Scarcelli's experiment is
straightforward.   Even if the source is a ``classical" chaotic
light, the ghost image is explained as a quantum two-photon
effect.  In a joint photodetection event of chaotic radiation, due to the 
superposition of two-photon amplitudes,  there exists 50\%
chance for a precise position-position correlation, i.e., if 
photon 1 is measured at $\vec{\rho}_1$, photon two can only be
measured at a unique certain position of $\vec{\rho}_2$.   
The following is a brief quantum mechanical calculation for the 
second-order correlation of chaotic radiation.  

In the quantum theory of photodetection \cite{Glauber},
the second-order correlation function is given in Eq.~(\ref{G2}).
What we need to do now is to find the state of the radiation and the 
operators of the field.
In the \emph{{photon counting}} regime, it is reasonable to model the 
thermal light in terms of \emph{{single photon states}}.  We have provided
a simple model of thermal state in Appendix B.  Similar to our 
earlier discussion, we will focus on the transverse 
correlation and assume the thermal radiation monochromatic:
\begin{eqnarray}\label{densitythermal}
\hat{\rho} \simeq |\,0\,\rangle\langle\,0\,| + \left\vert \epsilon\right\vert ^{4}
\sum_{\vec{\kappa}}\sum_{\vec{\kappa}'} 
\hat{a}^{\dagger}(\vec{\kappa}) \hat{a}^{\dagger}(\vec{\kappa}') \, | \,0 \, \rangle
\langle \, 0 \, | \, \hat{a}(\vec{\kappa}') \hat{a}(\vec{\kappa})
\end{eqnarray}
where $|\epsilon| \ll 1$. Basically we are modeling the light source as an 
incoherent statistical mixture of two photons with equal probability of
having any transverse momentum $\vec{\kappa}$ and $\vec{\kappa}'$.  

The transverse spatial part of the second order correlation function is thus:   
\begin{eqnarray}\label{G2-1}
G^{(2)}(\vec{\rho}_{1};\vec{\rho}_{2}) &= &
\sum_{\vec{\kappa},\vec{\kappa}'} \langle 1_{\vec{\kappa}}1_{\vec{\kappa}'}|
E_{1}^{(-)}(\vec{\rho}_{1})E_{2}^{(-)}(\vec{\rho}_{2})
E_{2}^{(+)}(\vec{\rho}_{2})E_{1}^{(+)}(\vec{\rho}_{1})
|1_{\vec{\kappa}}1_{\vec{\kappa}'}\rangle
\nonumber \\ &=& \sum_{\vec{\kappa},\vec{\kappa}'} \big{|} \langle 0| \,
E_{2}^{(+)}(\vec{\rho}_{2})E_{1}^{(+)}(\vec{\rho}_{1})
|1_{\vec{\kappa}}1_{\vec{\kappa}'}\rangle \, \big{|}^{2}.
\end{eqnarray}
where $\vec{\rho}_{j}$ is the transverse coordinate of the $j^{th}$
detector. The transverse part of the electric field operator can
be written as follows:
\begin{eqnarray}\label{E12}
E^{(+)}_{j}(\vec{\rho}_{j})\propto\sum_{\vec{\kappa}} \,
g_{j}(\vec{\rho}_{j};\vec{\kappa}) \, \hat{a}(\vec{\kappa})
\end{eqnarray}
where $\hat{a}(\vec{\kappa})$ is the annihilation operator for the mode
corresponding to $\vec{\kappa}$ and $g_{j}(\vec{\rho}_{j};\vec{\kappa})$ is the
Green's function associated to the propagation of the field from
the source to the $j^{th}$ detector \cite{Rubin}.

Substituting the field operators into Eq.~(\ref{G2-1}) we obtain:
\begin{eqnarray}\label{G2-2}
G^{(2)}(\vec{\rho}_{1};\vec{\rho}_{2}) 
=\sum_{\vec{\kappa},\vec{\kappa}'}
\big{|} g_{2}(\vec{\rho}_{2},\vec{\kappa})g_{1}(\vec{\rho}_{1},\vec{\kappa}')+
g_{2}(\vec{\rho}_{2},\vec{\kappa}')g_{1}(\vec{\rho}_{1},\vec{\kappa}) \big{|}^{2}.
\end{eqnarray}
This expression represents the key result toward the understanding
of the phenomenon. In fact, it expresses an interference between
two alternatives, different yet equivalent, which lead to a joint
photodetection. The interference phenomenon is not, as in
classical optics, due to the superposition of electromagnetic
fields at a local point of space-time. It is due to the superposition
of $g_{2}(\vec{\rho}_{2},\vec{\kappa})g_{1}(\vec{\rho}_{1},\vec{\kappa}')$ and
$g_{2}(\vec{\rho}_{2},\vec{\kappa}')g_{1}(\vec{\rho}_{1},\vec{\kappa})$, the
so-called two-photon amplitudes, non-classical entities that
involve both arms of the optical setup as well as two distant photodetection
events. Eq.~(\ref{G2-2}) can be further evaluated into the form of
\begin{eqnarray}\label{G2-3-2}
G^{(2)}(\vec{\rho}_{1};\vec{\rho}_{2}) & \propto&\sum_{\vec{\kappa}}
|g_{1}(\vec{\rho}_{1},\vec{\kappa})|^{2}\sum_{\vec{\kappa}'}
|g_{2}(\vec{\rho}_{2},\vec{\kappa}')|^{2}
 + |\sum_{\vec{\kappa}}g^{*}_{1}(\vec{\rho}_{1},\vec{\kappa})
g_{2}(\vec{\rho}_{2},\vec{\kappa})|^{2}
\nonumber \\ &=& G^{(1)}_{11}(\vec{\rho}_{1})G^{(1)}_{22}(\vec{\rho}_{2})+
|G^{(1)}_{12}(\vec{\rho}_{1};\vec{\rho}_{\rho})|^{2}.
\end{eqnarray}
In Eq.~(\ref{G2-3-2}), we have linked $G^{(2)}$ with the 
first order correlation functions of $G^{(1)}_{ij}$.
The first term in Eq.~(\ref{G2-3-2}) is the product of the mean
intensities measured by the two detectors.  The second term, 
which corresponds to the ``intensity fluctuation correlation" in 
Eq.~(\ref{Intensity-1}), is nothing but the two-photon interference term.
The superposition takes place between quantities
$g_{2}(\vec{\rho}_{2},\vec{\kappa})g_{1}(\vec{\rho}_{1},\vec{\kappa}')$ and
$g_{2}(\vec{\rho}_{2},\vec{\kappa}')g_{1}(\vec{\rho}_{1},\vec{\kappa})$ in
Eq.~(\ref{G2-2}), namely the two-photon amplitudes.   

Now, we calculate the
$|G^{(1)}_{12}(\vec{\rho}_{1};\vec{\rho}_{2})|^{2}$ term of Eq.~(\ref{G2-3-2}). 
Following the experimental setup of
Fig.~\ref{fig:HBT-Fig1} (b), the Green's function of free-propagation 
can be written as:
\begin{eqnarray*}
g_{1}(\vec{\rho}_{1};\vec{\kappa}) = \int d \vec{\rho}_s \, 
\Big\{\frac{-i \omega}{2 \pi c} \, 
\frac{e^{i \frac{\omega}{c} d_A} }{d_A}\, e^{i \frac{\omega}{2 c d_A} 
|\vec{\rho}_{1}-\vec{\rho}_{s}|^2} \Big\} \, 
e^{-i \vec{\kappa} \cdot \vec{\rho}_s} \\
g_{2}(\vec{\rho}_{2};\vec{\kappa}) =  \int d \vec{\rho'_s} \, 
\Big\{\frac{-i \omega}{2 \pi c} \, 
\frac{e^{i \frac{\omega}{c} d_B} }{d_B}\, e^{i \frac{\omega}{2 c d_B} 
|\vec{\rho}_{2}-\vec{\rho'_s}|^2} \Big\} \, 
e^{-i \vec{\kappa} \cdot \vec{\rho'_s}} 
\end{eqnarray*}
where $\vec{\rho}_s$ is the transverse vector on the source plane. 
We have therefore propagated the field from the source to the $\vec{\rho}_{1}$
plane and $\vec{\rho}_2$ plane in arm A and arm B, respectively. 

Substituting $g^{*}_{1}(\vec{\rho}_{1},\vec{\kappa})$ and 
$g_{2}(\vec{\rho}_{2},\vec{\kappa})$ into 
$G^{(1)}_{12}(\vec{\rho}_{1};\vec{\rho}_{2})$ and completing the integral on 
$d\vec{\kappa}$, we obtain
\begin{eqnarray}\label{G12-Integral}
G^{(1)}_{12}(\vec{\rho}_{1};\vec{\rho}_{2}) &=& \int d\vec{\kappa} \,\, 
g^{*}_{1}(\vec{\rho}_{1},\vec{\kappa}) \, 
g_{2}(\vec{\rho}_{2},\vec{\kappa}) \nonumber \\
&\propto& \int d \vec{\rho}_s \, e^{-i \frac{\omega}{c} d_A} \, 
e^{-i \frac{\omega}{2 c d_A} |\vec{\rho}_{1}-\vec{\rho}_{s}|^2} \, 
e^{i \frac{\omega}{c} d_B} \, e^{i \frac{\omega}{2 c d_B} |\vec{\rho}_{2}-\vec{\rho}_{s}|^2}
\end{eqnarray}
If we chose the distances from the source to the two detectors equal
($d_{A}=d_{B}$), the integral of $d \vec{\rho}_s$ in Eq.~(\ref{G12-Integral}) 
yields a point-point correlation, $\delta (\vec{\rho}_{1} - \vec{\rho}_{2})$, 
between the $\vec{\rho}_{1}$  plane and 
the $\vec{\rho}_{2}$ plane, while taking
an infinite size of the source.  Thus, after the integration of the
bucket detector, the joint-detection counting rate is calculated as:
\begin{eqnarray}\label{imagemir}
R_c (\vec{\rho}_{2}) \propto \int
d\vec{\rho}_{1} \, |A(\vec{\rho}_{1}) \, \delta(\vec{\rho}_{1}-\vec{\rho}_{2})|^{2}
= |A(\vec{\rho}_{2})|^{2}
\end{eqnarray}
where $A(\vec{\rho}_{1})$ is the aperture function of the object plane.

Thus we have successfully explained the experimental observation
in terms of two-photon interference.  

As we know $G^{(1)}_{12}$ has a counterpart of $\Gamma^{(1)}_{12}$ in
classical optics, namely the first-order coherence function.  Can we use
classical interference to explain this phenomenon?
In fact, the theory of statistical correlation of intensity fluctuations is not 
a ``must" for explaining the second-order correlation of thermal radiation. 
Examining $G^{(2)} = G^{(1)}_{11}G^{(1)}_{22} +
|G^{(1)}_{12}|^{2}$ of thermal light, it is easy to find that the intensity fluctuation 
correlation corresponds to 
the second term of $|G^{(1)}_{12}|^{2}$. $G^{(1)}_{12}$ characterizes 
the first-order coherence, i.e., the ability of observing first-order interference, 
of the fields $E(\mathbf{r}_1, t_1)$ and $E(\mathbf{r}_2, t_2)$. 
Why don't we explain HBT in terms of classical 
interference, instead of the intensity fluctuation correlation?  
As we usually say: ``people are always smarter than theory."  
This time, however, people have been too smart (we lost 50 years)!  There is a reason 
not to do that, because $E(\mathbf{r}_1, t_1)$ and $E(\mathbf{r}_2, t_2)$
would not interfere with each other in the case of HBT: 
$E(\mathbf{r}_1, t_1)$ and $E(\mathbf{r}_2, t_2)$ are measured by two 
photodetectors at distance.  Maxwell's wave equation is a linear differential 
equation. If $E(\mathbf{r}_1, t_1)$ and $E(\mathbf{r}_2, t_2)$ are solutions 
of the wave equation then the resulting field of $E(\mathbf{r}, t) = 
E(\mathbf{r}_1, t_1) + E(\mathbf{r}_2, t_2)$ is also a solution.  However, 
to claim a solution of the wave equation or interference, $E(\mathbf{r}_1, t_1)$ 
and $E(\mathbf{r}_2, t_2)$ have to be physically 
superposed at a space-time point, as we usually do in an interferometer.  
For example, in the YoungÕs double slit experiment, $E(\mathbf{r}_1, t_1)$ and 
$E(\mathbf{r}_2, t_2)$  are the fields at the double slits in early times relative to the 
time of photodetection. The resulting field of $E(\mathbf{r}, t) = 
E(\mathbf{r}_1, t_1) + E(\mathbf{r}_2, t_2)$ is physically added at the space-time 
point of the photodetection. In HBT and Scarcelli's experiment, however, 
the situation is different. $E(\mathbf{r}_1, t_1)$ 
and $E(\mathbf{r}_2, t_2)$ are never brought together; further more, 
$E(\mathbf{r}_1, t_1)$ and $E(\mathbf{r}_2, t_2)$ are measured by two distant 
photodetectors!  What kind of interference is this?  

It is interference, but a different type.  Under the framework of Glauber's photodetection 
theory, we have provided a two-photon interference picture. It is the interference 
between two-photon amplitudes.  Unfortunately, this concept has no counterpart in 
classical electromagnetic theory of light, 
unless one makes the theory ÒnonlocalÓ by assuming $E(\mathbf{r}_1, t_1)$ 
and $E(\mathbf{r}_2, t_2)$ are addable nonlocally through the measurement of two 
photodetectors, or makes a concept such as the interference between 
[field A goes to $D_1$ $\times$ field B goes to $D_2$] and [field B goes to $D_1$ 
$\times$ field A goes to $D_2$], which does not make any sense under the framework 
of Maxwell electromagnetic wave theory of light.  

To conclude this section: the two-photon imaging experiment of 
Scarcelli \emph{et al.} has demonstrated that the ghost imaging of chaotic
light is an interference phenomenon involving the superposition
between indistinguishable two-photon alternatives, 
rather than statistical correlation of intensity fluctuations. 
The two-photon correlations are \textit{observed in} the intensity
fluctuations, however, they are \textit{not caused by} the
statistical correlation of the intensity fluctuations.

\vspace{3mm}

\section*{A statement from the author}

\hspace{6.5mm}This article was originally prepared as lecture notes for my students. 
The lecture notes were turned into a review paper at the end of 2006.  
The main purpose of this article is to clarify the fundamental concepts of quantum 
imaging and to provide the basic knowledge of physics and necessary tools of 
mathematics, which are involved in quantum imaging, for the general physics and 
engineering community. The debate regarding the quantum or classical nature of 
quantum imaging is still ongoing and will probably endure for a while.  In a sense, 
this is a debate that started 50 years ago since the discovery of HBT.  In this article, 
we have emphasized the two-photon coherence nature of the quantum imaging 
phenomena.  As usual, there are always different opinions \cite{comm}. Classical 
formalisms do exist, such as Eq.~(\ref{HBT-Classical}); such formalism, however, 
is far from capturing the essence of the phenomenon and may mislead us from 
the truth, as it has happened to the two-photon physics of thermal light.  The 
interpretation in terms of two-photon coherence elegantly explains all of the 
features of the experiments carried out with both ``quantum" and ``classical" 
sources.  For this reason, two-photon coherence seems more suitable for 
explaining the physics of quantum imaging.

\vspace*{5mm}

\section*{Appendix A: Fresnel diffraction-propagation}
\def\theequation{$A-$\arabic{equation}}
\setcounter{equation}{0}
\def\thefigure{$A-$\arabic{figure}}
\setcounter{figure}{0}

\begin{figure}[hbt]
 \centering
    \includegraphics[width=75mm]{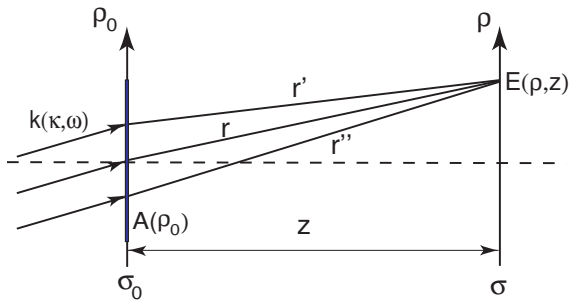}
     \parbox{14cm}{\caption{Schematic of free-space Fresnel propagation.  
    The complex aperture function $\tilde{A}(\vec{\rho}_0)$ 
    is composed by a real function $A(\vec{\rho}_0)$ and a phase 
    $e^{-i \vec{\kappa} \cdot \vec{\rho}_0}$ associated with each of the transverse 
    wavevector $\vec{\kappa}$ on the plane of $\sigma_0$.  Only one mode of 
    wavevector $\mathbf{k}(\vec{\kappa}, \omega)$ is shown in the figure.}
    \label{fig:Fresnel}}
\end{figure}

\hspace{6.5mm}In Fig.~\ref{fig:Fresnel}, the field is freely propagated from plane 
$\sigma_0$ to plane $\sigma$.  
For imaging applications, it is convenient to describe such a propagation in the 
form of Eq.~(\ref{e-g}).  In this Appendix, we evaluate the 
$g(\vec{\kappa}, \omega; \vec{\rho}, z)$, namely the Green's function or
the field transfer function, for free-space Fresnel propagation.  

According to the Huygens-Fresnel principle, the field at space-time point
$(\vec{\rho}, z, t)$ is the result of a superposition among these spherical 
secondary wavelets originated from each point on the $\sigma_0$ plane,
see Fig.~\ref{fig:Fresnel},
\begin{eqnarray}
E(\vec{\rho}, z, t) = \int d\vec{\kappa} \, d\omega \,
\tilde{E}(\vec{\kappa}, \omega) \int d\vec{\rho}_0 \, 
\frac{\tilde{A}(\vec{\rho}_0)}{r'} \, e^{-i (\omega t - k r')}
\end{eqnarray}
where $\tilde{A}(\vec{\rho}_0)$ is the complex aperture function and
we have taken $z_0 = 0$ and $t_0 = 0$ as usual.  We have also assumed
a common complex amplitude $\tilde{E}(\vec{\kappa}, \omega)$ for all point
sources of the secondary wavelets on the plane of $\sigma_0$.  
In a paraxial approximation,
we take the first-order expansion of $r'$ in terms of $z$ and $\vec{\rho}$ 
$$
r' = \sqrt{z^2 + |\vec{\rho} - \vec{\rho}_0|^2} \simeq z(1 + 
\frac{|\vec{\rho} - \vec{\rho}_0|^2}{2 z^2}).
$$
$E(\vec{\rho}, z, t)$ is thus approximated as
\begin{eqnarray}
E(\vec{\rho}, z, t) 
\simeq  \int d\vec{\kappa} \, d\omega \, 
\tilde{E}(\vec{\kappa}, \omega) \int d\vec{\rho}_0 \, \frac{\tilde{A}(\vec{\rho}_0)}{z}
\, e^{i \frac{\omega}{c} z} \, 
e^{i \frac{\omega}{2 c z} |\vec{\rho} - \vec{\rho}_0|^2} e^{-i \omega t}
\end{eqnarray}
where $e^{i \frac{\omega}{2 c z} |\vec{\rho} - \vec{\rho}_0|^2}$ is named as
the Fresnel phase factor.

Assuming the complex aperture function is composed by a real function 
$A(\vec{\rho}_0)$ and a phase $e^{-i \vec{\kappa} \cdot \vec{\rho}_0}$ 
associated with the transverse
wavevector and the transverse coordinate on the plane of $\sigma_0$, 
which is reasonable for the setup of Fig.~\ref{fig:Fresnel},
$E(\vec{\rho}, z, t)$ can be written in the following form
\begin{eqnarray}
E(\vec{\rho}, z, t) =  \int d\vec{\kappa} \, d\omega \,\tilde{E}(\vec{\kappa}, \omega)  \, 
e^{-i \omega t} 
\int d\vec{\rho}_0 \, A(\vec{\rho}_0) \, e^{i \vec{\kappa} \cdot \vec{\rho}_0} \, 
\frac{e^{i \frac{\omega}{c} z}}{z} \, 
e^{i \frac{\omega}{2 c z} |\vec{\rho} - \vec{\rho}_0|^2}.
\end{eqnarray}
The Green's function for free-space Fresnel propagation is thus
\begin{eqnarray}\label{gg-final}
g(\vec{\kappa}, \omega; \vec{\rho}, z) 
=  \frac{-i \omega}{ 2 \pi c} \ \frac{e^{i \frac{\omega}{c} z}}{z}
\int_{\sigma_0} \, d\vec{\rho}_0 \, \tilde{A}(\vec{\rho}_0) \, 
e^{i \frac{\omega}{2 c z} |\vec{\rho} - \vec{\rho}_0|^2}
\end{eqnarray}
where $- i \omega / 2 \pi c$ is a normalization constant.

\section*{Appendix B: Quantum state of thermal light}
\def\theequation{$B-$\arabic{equation}}
\setcounter{equation}{0}
\def\thefigure{$B-$\arabic{figure}}
\setcounter{figure}{0}

\hspace{6.5mm}We assume a large number
of atoms that are ready for two-level atomic transition. At most times, the
atoms are in their ground state. There is, however, a small chance for each
atom to be excited to a higher energy level and later release a photon during
an atomic transition from the higher energy level $E_{2}$ ($\Delta E_{2}
\neq0$) back to the ground state $E_{1}$. It is reasonable to assume that each
atomic transition excites the field into the following state:
\begin{eqnarray}
|\, \Psi\, \rangle  \simeq|\, 0 \, \rangle+ \epsilon\,
\sum_{\mathbf{k},s} \, f(\mathbf{k},s) \, \hat{a}^{\dag}_{\mathbf{k},s}\,|\, 0
\, \rangle
\end{eqnarray}
where $|\epsilon| \ll1$ is the probability amplitude for the atomic
transition. Within the atomic transition, $f(\mathbf{k},s) = \langle\,
\Psi_{\mathbf{k},s} \, |\, \Psi\, \rangle$ is the probability amplitude for
the radiation field to be in the single-photon state of wave number
${\mathbf{k}}$ and polarization $s$: $|\, \Psi_{\mathbf{k},s} \, \rangle=
|\,1_{\mathbf{k},s} \, \rangle= \hat{a}^{\dag}_{\mathbf{k},s}\,|\, 0 \,
\rangle$.

For this simplified two-level system, the density matrix that characterizes
the state of the radiation field excited by a large number of possible atomic
transitions is thus
\begin{align}\label{B-2}
\hat{\rho}  &  =\prod_{t_{0j}}\,\Big\{|\,0\,\rangle+\epsilon\sum
_{\mathbf{k},s}\,f(\mathbf{k},s)\,e^{-i\omega t_{0j}}\,\hat{a}_{\mathbf{k}%
,s}^{\dag}\,|\,0\,\rangle\,\Big\}\\ \nonumber
&  \times\,\prod_{t_{0k}}\,\Big\{\langle\,0\,|\,+\epsilon^{\ast}%
\sum_{\mathbf{k}^{\prime},s^{\prime}}\,f(\mathbf{k}^{\prime},s^{\prime
})\,e^{i\omega^{\prime}t_{0k}}\,\langle\,0\,|\hat{a}_{\mathbf{k}^{\prime
},s^{\prime}}\,\Big\}\\ \nonumber
&  \simeq\Big\{|\,0\,\rangle+\epsilon\lbrack \sum_{t_{oj}}\sum_{\mathbf{k}%
,s}\,f(\mathbf{k},s)\,e^{-i\omega t_{0j}}\,\hat{a}_{\mathbf{k},s}^{\dag
}\,|\,0\,\rangle] + \epsilon^2[...] \Big\} \\ \nonumber
&  \times\,\,\Big\{\langle\,0\,|\,+\epsilon^{\ast}[\sum_{t_{ok}}\sum_{\mathbf{k}^{\prime
},s^{\prime}}\,f(\mathbf{k}^{\prime},s^{\prime})\,e^{i\omega^{\prime}t_{0k}%
}\,\langle\,0\,|\hat{a}_{\mathbf{k}^{\prime},s^{\prime}}] + \epsilon^{\ast 2}[...] \Big\}
\end{align}
where $e^{-i\omega t_{0j}}$ is a random phase factor associated with the 
jth atomic transition. Since $\left\vert\epsilon\right\vert \ll1$, it is a good approximation 
to keep the necessary lower-order terms of $\epsilon$ in Eq.(\ref{B-2}).  
After summing over $t_{0j}$ ($t_{0k}$) by taking all its
possible values, approximately, we have 
\begin{align}
\hat{\rho} &\simeq|\,0\,\rangle\langle\,0\,|+ \left\vert \epsilon\right\vert
^{2}\sum_{\mathbf{k,s}}
\left\vert f(\mathbf{k,s})\right\vert ^{2}
|\,1_{\mathbf{k,s}}\,\rangle
\langle\,1_{\mathbf{k,s}}\,|,\nonumber \\
& + \left\vert \epsilon\right\vert
^{4}\sum_{\mathbf{k,s}}\sum_{\mathbf{k}^{\prime}, s^{\prime}}
\left\vert f(\mathbf{k,s})\right\vert ^{2} \left\vert f(\mathbf{k}^{\prime},s^{\prime})\right\vert ^{2}
|\,1_{\mathbf{k,s}}1_{\mathbf{k}^{\prime},s^{\prime}}\,\rangle
\langle\,1_{\mathbf{k},s}1_{\mathbf{k}^{\prime},s^{\prime}}\,|.
\end{align}


\begin{thebibliography}{1}
\bibitem{aberration} Even if assuming a perfect lens without
chromatic aberration, Fresnel diffraction is wavelength dependent.
Hence, large broadband ($\Delta \omega \sim \infty$) would result in 
blurred images in classical imaging.  Surprisingly, the situation is
different in quantum imaging: no aberration blurring.
\bibitem{Speckle} A popular perception views the two-photon phenomenon 
as the correlation between identical copies of ``speckles" across the two 
light beams.  Is this idea incorrect?  One may simply split an attenuated 
(to single-photon's level) or scattered laser beam into two by a beamsplitter 
to observe identical ``speckles" either temporally or spatially on the two 
beams and yet, in this situation, the second-order correlation is a constant. 
On the other hand, for a smooth bright thermal light beam, which is achievable
easily and classically, one may not be able to 
observe any ``speckles", the HBT two-photon correlation is still observable.
\end{thebibliography}
\end{document}